\documentclass{ws-p8-50x6-00}
\usepackage{epsfig}
\def\gapx{\lower 2pt
\hbox{$\buildrel>\over{\scriptstyle{\sim}}$}}
\def\lapx{\lower 2pt \hbox{$\buildrel<\over{\scriptstyle{\sim}}$}}
\def\4he{$^4$He}

\def\br{{\bf r}}

\def\beq{\begin{equation}}
\def\eeq{\end{equation}}
\def\bea{\begin{eqnarray}}
\def\eea{\end{eqnarray}}

\def\bp{{\bf p}}
\def\bv{{\bf v}}
\def\tn{\tilde n}

\def\be{\bbox{\eta}}

\begin{document}

\title{Condensate oscillations, kinetic equations and two-fluid 
hydrodynamics in a Bose gas}

\author{Allan Griffin}

\address{Department of Physics, University of Toronto, Toronto, 
Ontario, Canada M5S 1A7\\E-mail: griffin@physics.utoronto.ca}

\maketitle

\abstracts{ }

\section{Introduction}
\label{sec:Introduction}

Trapped Bose-condensed atomic gases~\cite{Dalgiorpit,Ingstrwie} are 
remarkable because, in spite of the fact that these are very dilute 
systems, they exhibit robust coherent dynamic behaviour when 
perturbed.  These quantum ``wisps of matter'' are a new phase of 
highly coherent matter.  While binary collisions are very infrequent, 
the coherent mean field associated with the Bose condensate ensures 
that interactions play a crucial role in determining the collective 
response of these superfluid gases.

In our discussion of the theory of collective oscillations of atomic 
condensates, which is the main theme of these lectures, the 
macroscopic Bose wavefunction $\Phi({\bf r}, t)$ will play a central 
role.  This wavefunction is the BEC order parameter.  The initial 
attempts at defining this order parameter began with the pioneering 
work of Fritz London~\cite{Flondon} in 1938, was further developed by 
Bogoliubov~\cite{Nnbog} in 1947 and finally formalized in the general 
quantum field theoretic formalism of Beliaev~\cite{Bel2} in 1957.  
The first extension of these ideas to inhomogeneous Bose condensates 
was by Pitaevskii and, independently, by Gross in 1961, which led to 
the now famous Gross-Pitaevskii (GP) equation of motion for 
$\Phi({\bf r}, t)$.  Most of this early work was limited to $T=0$ 
where, in a dilute Bose gas, one can assume all of the atoms are in 
the condensate.

In Section~\ref{sec:Pure}, I will first review the dynamics of a pure 
condensate at $T=0,$ based on solving the linearized GP equation.  
This limit is especially appealing since one can ignore all the 
complications which arise from the presence of non-condensate atoms 
(the thermal cloud).  We will discuss the normal mode solutions of 
the $T=0$ GP equation of motion using the ``quantum hydrodynamic'' 
formalism, which works in terms of the local condensate density 
$n_c({\bf r}, t)$ and superfluid velocity $\bv_c({\bf r}, t)$.  
Within the Thomas-Fermi approximation, Stringari~\cite{Sstringari} 
has shown that the equations of motion for these two variables can be 
combined to give a wave equation for oscillations of the condensate.  

In Section~\ref{sec:temperatures}, we derive a generalized form of 
the GP equation for $\Phi({\bf r}, t)$ which is valid at finite 
temperatures.~\cite{Zarnikgrif} It involves terms which are coupled 
to the non-condensate component (the thermal cloud) and thus its 
solution in general depends on knowing the equations of motion for 
the dynamics of the non-condensate.  In the present lectures, we will 
restrict ourselves to finite temperatures where the non-condensate 
can be described by a quantum kinetic equation for the 
single-particle distribution function $f({\bf p, \bf r}, t).$  In 
this Boltzmann equation, the relatively high energy non-condensate 
atoms are simply free atoms moving in a self-consistent Hartree-Fock 
mean field. 

A unique feature of a Bose-condensed gas is that the kinetic equation 
for $f({\bf p},{\bf r}, t)$ involves a collision integral (denoted 
by  $C_{12}[f, \Phi])$ describing collisions between condensate and 
non-condensate atoms.
The generalized GP equation for $\Phi(\br, t)$ also has a term which 
is related to the $C_{12}$ collisions.  This gives rise to damping of 
condensate fluctuations. In Section~\ref{sec:temperatures}, we derive 
a finite $T$ Stringari wave equation~\cite{Wilgrif} with damping 
using the static Popov approximation.  This means the thermal cloud 
is treated as always being in static thermal equilibrium, with 
$f({\bf p, \bf r}, t) = f_0({\bf p, \bf r})$ being given by the 
equilibrium Bose distribution.

In Section~\ref{sec:coupled}, we turn to a detailed treatment of the 
dynamics of the coupled condensate and non-condensate components, 
starting from the finite $T$ generalized GP equation for $\Phi({\bf 
r}, t)$ and our kinetic equation for $f({\bf p, r}, t)$.  We derive a 
new set of two-fluid hydrodynamic equations, following the recent 
work by Zaremba, Nikuni and Griffin.~\cite{Zarnikgrif}  This 
derivation assumes the non-condensate is in {\it local} hydrodynamic 
equilibrium, induced by rapid collisions between the atoms in the 
thermal cloud.  As a result, the non-condensate is completely 
described in terms of the non-condensate density ${\tilde n}({\bf r}, 
t),$ velocity ${\bf v}_n({\bf r}, t)$ and local pressure ${\tilde 
P}({\bf r}, t)$.  Both the condensate and non-condensate exhibit 
coupled coherent oscillations at the {\em same} frequency.

In Section~\ref{sec:fluid}, we review the famous two-fluid 
hydrodynamic equations first derived by Landau in 1941 and which form 
the basis of our understanding of the hydrodynamic behaviour of 
superfluid $^4$He.~\cite{Land,Khalatnikov}  We discuss first sound 
and second sound and consider how they differ  in superfluid $^4$He 
and in a  uniform Bose gas.  We also point out the appearance of a 
new hydrodynamic zero frequency mode which appears in our generalized 
two-fluid equations.  This mode is associated with the relaxation 
time for the condensate and non-condensate atoms to come into 
``diffusive'' equilibrium with each other.  Inclusion of damping due 
to thermal conductivity of the thermal cloud shows that this mode is 
the analogue of the well-known thermal diffusion mode in a classical 
gas.

We close this Introduction with some general references which may be 
useful to the reader.  The 1998 Varenna Summer School lectures on 
BEC~\cite{Ingstrwie} has excellent articles on recent research on BEC 
in atomic gases, including very detailed reviews of the experimental 
work carried out by the JILA~\cite{cornell} and MIT~\cite{ketterle} 
groups.  For a general introduction to the theory of trapped Bose 
gases, we recommend the Reviews of Modern Physics~\cite{Dalgiorpit} 
article by a leading theoretical group at the University of Trento.  
This authoritative review concentrates on thermodynamic properties 
and the collisionless dynamics.  A recent long paper by Zaremba, 
Nikuni and the author~\cite{Zarnikgrif} discusses the details of the 
derivation of the coupled hydrodynamic equations for the condensate 
and the thermal cloud.  The present lectures may be viewed as an 
introduction to this paper, with more emphasis on the general 
structure of the derivation and the implications of this new set of 
superfluid equations.  	

\section{Pure condensate dynamics (at $T=0$)}
\label{sec:Pure}
This section will be largely a review of standard 
material~\cite{Dalgiorpit} but will provide the starting point for 
generalizations in the following sections.  As we have mentioned, at  
$T=0$, one can assume that all atoms in a dilute Bose gas are 
described by the macroscopic wavefunction 
 $\Phi({\bf r}, t)$.  This obeys the time-dependent Hartree equation 
of motion first written down 40 years ago by Pitaevskii and Gross,
\begin{equation}
i\hbar{\partial\Phi\over\partial t}({\bf r}, t) = 
\left[-{\hbar^2\nabla^2\over 2m}+V_{ex}({\bf r}) + V_{H}({\bf r}, 
t)\right]\Phi({\bf r}, t).
\label{eq:dynamics1}
\end{equation}
Here the trap harmonic potential is
\begin{equation}
V_{ex}({\bf r}) = {1\over 2}\, m\omega^2_0 r^2 \ \ \ \ \ 
\mbox{(isotropic)}
\label{eq:dynamics2}
\end{equation}
and the self-consistent condensate Hartree potential is

\begin{equation}
V_{H}({\bf r}, t) = \int d{\bf r}^\prime v({\bf r}-{\bf r}^\prime) 
n_c({\bf r}, t) = gn_c({\bf r}, t).\label{eq:dynamics3}\end{equation}
As usual, since we are interested in extremely low energy atoms, we 
use the $s-$wave approximation and approximate the interatomic 
potential by the pseudopotential
\bea
v({\bf r}-{\bf r}^\prime) &=& 
{4\pi a\hbar^2\over m}\delta({\bf r}-{\bf r}^\prime) 
 \nonumber \\[4pt]
&=&g\delta({\bf r}-{\bf r}^\prime),
\label{eq:dynamics4}
\eea
where $a$ is the correct $s-$wave scattering length.  For further 
discussion of the effective interaction to use in the 
Gross-Pitaevskii (GP) equation, we refer to the lectures by  Burnett 
in this volume.  The condensate density is given by $n_c({\bf r}, t) 
= |\Phi({\bf r}, t)|^2$ and hence (\ref{eq:dynamics1}) reduces to a 
NLSE for $\Phi({\bf r}, t),$ namely
\begin{equation}
i\hbar{\partial\Phi({\bf r}, t)\over\partial t} = 
\left[-{\hbar^2\nabla^2_r\over 2m}+V_{ex}({\bf r}) +g|\Phi({\bf r}, 
t)|^2\right]\Phi({\bf r}, t).
\label{eq:dynamics5}
\end{equation}
Since all atoms are in the identical quantum state, there is no 
exchange mean field in (\ref{eq:dynamics1}).

The $T=0$ GP equation in (\ref{eq:dynamics5}) has been the subject of 
literally hundreds of papers since the discovery of BEC in 
laser-cooled trapped atomic gases - and the equation appears in 
almost every chapter in this book.  As discussed in recent 
reviews~\cite{Dalgiorpit,Ingstrwie}, it gives an excellent 
quantitative description of both the static and dynamic (linear and 
non-linear) behaviour in trapped Bose gases below about $T\,\lapx\, 
0.5T_{BEC}.$  The accuracy can be of the order of a few percent, 
which is as much as one can expect since the non-condensate fraction 
of atoms at $T=0$ is also estimated to be about $1\%$ or 
so.~\cite{Hutzargrif}  The $T=0$ GP equation has been extended to 
deal with two-component Bose gases (involving two different atomic 
hyperfine states) and the effect of perturbations related to laser 
and rf fields (used to manipulate the Bose condensates) is easily 
incorporated. For further discussion and references, see the lectures 
by Ballagh in this book.

Our main purpose in this section is to use the GP equation 
(\ref{eq:dynamics5}) to discuss collective oscillations of a pure 
condensate.  In Section~\ref{sec:temperatures}, we will discuss the 
extension of this equation at finite $T$ to deal with the effect of 
non-condensate atoms.  We first briefly review the static equilibrium 
solution $\Phi_0({\bf r})$ of the GP equation, given by 
\begin{equation}
\left<{\hat\psi}({\bf r}, t)\right>\equiv\Phi({\bf r}, t)
=\Phi_0({\bf r})e^{-i\mu t/\hbar},
\label{eq:dynamics6}
\end{equation}
where $\mu$ is the chemical potential of the condensate.  The physics 
behind this can be seen from
\bea
 \left<N-1|{\hat\psi}({\bf r}, t)|N\right>
&=& e^{iE_{N-1} t/\hbar} \left<N-1|{\hat\psi}({\bf 
r})|N\right>e^{-iE_N t/\hbar} \nonumber\\
&=&\left<N-1|\sqrt{N}|N-1\right>e^{-i(E_N-E_{N-1})t/\hbar}
\nonumber\\
&=& \sqrt{N} e^{-i\mu t/\hbar}.
\label {eq:dynamics7}
\eea
Using (\ref{eq:dynamics6}) in (\ref{eq:dynamics5}) gives
\begin{equation}
\mu_{c0}\Phi_0({\bf r})=\left[-{\hbar^2\nabla^2\over 2m} +V_{ex}({\bf 
r})+g|\Phi_0({\bf r})|^2\right]\Phi_0({\bf r}).\label{eq:dynamics8}
\end{equation}
Assuming a vortex-free ground state, this GP equation for the static 
condensate wavefunction  $\Phi_0({\bf r}) = \sqrt{n_{c0}({\bf r})}$ 
leads to the following expression for the equilibrium condensate 
chemical potential
\begin{equation}
\mu_{c0} = -{\hbar^2\nabla^2\sqrt{n_{c0}({\bf r})}\over 
2m\sqrt{n_{c0}({\bf r})}} +V_{ex}({\bf r}) +gn_{c0}({\bf 
r}).\label{eq:dynamics9}
\end{equation}

A standard approximation in solving (\ref{eq:dynamics9}) is to ignore 
the kinetic energy associated with the condensate amplitude 
$\sqrt{n_c},$ ie, neglect the $-{\hbar^2\nabla^2\over 2m}$ term.  In 
this ``Thomas-Fermi''  (TF) approximation,  the static GP  equation 
for $\Phi_0({\bf r})$ reduces to~\cite{Dalgiorpit}
\begin{equation}
\left[V_{ex}({\bf r}) + g|\Phi_0({\bf 
r})|^2\right]=\mu_{c0},\label{eq:dynamics10}
\end{equation}
which is easily inverted to give the condensate density profile
\bea
n_{c0}({\bf r}) &=& {1\over g}[\mu_{c0} - V_{ex}({\bf r})] 
\nonumber \\
&=& {1\over g}\left[\mu_{c0} - {1\over 2} 
m\omega^2_0r^2\right].\label{eq:dynamics11}
\eea
Clearly in the TF approximation, the size of the condensate is 
$R_{TF}$, where 
\begin{equation}
\mu_{c0} = {1\over 2} m\omega^2_0 R^2_{TF}. 
\label{eq:dynamics12}
\end{equation}
One finds $\mu_{c0}$ from the condition $\int d{\bf r}n_{c0}({\bf 
r})= N = N_c,$ which gives
\begin{equation}
\mu_{c0} = {\hbar\omega_0\over 2} \left[15{N_c a\over 
a_{HO}}\right]^{2/5}; \ \ a_{HO} \equiv (\hbar/m\omega_0)^{1/2}. 
\label{eq:dynamics13}
\end{equation}
We recall that the oscillator length $a_{HO}$ is the size of the 
ground state Gaussian wavefunction  of an atom in a parabolic 
potential.  Combining (\ref{eq:dynamics12}) and (\ref{eq:dynamics13}) 
gives
\bea
R_{TF} &=& a_{HO}\left(15 {N_ca\over a_{HO}}\right)^{1/5} \nonumber \\
& \gg & a_{HO}, \  \mbox{if} \ {N_ca\over a_{HO}}\gg 1. 
\label{eq:dynamics14}
\eea
The TF approximation (\ref{eq:dynamics11}) for $n_{c0}({\bf r})$ is 
very good for large $N_c$, except for a small region near the edge of 
condensate $(r\simeq R_{TF}).$  In practice, for typical values of 
$a_{HO}$ and $a$, one finds the TF approximation is very good for 
$N_c\,\gapx\, 10^4$ atoms.~\cite{Dalgiorpit}

To discuss condensate fluctuations around the static equilibrium 
value of $\Phi_0({\bf r})$, we first reformulate the time-dependent 
GP equation in terms of the condensate density and phase variables
\begin{equation}
\Phi({\bf r}, t) = \sqrt{n_c({\bf r}, t)}e^{i\theta({\bf r}, t)} . 
\label{eq:dynamics15}
\end{equation}
Inserting this into the GP equation (\ref{eq:dynamics5}), gives
\bea
i\hbar{\partial\sqrt{n_c}\over\partial t} 
-\sqrt{n_c}{\hbar\partial\theta\over\partial t} = \mu_c({\bf r}, t) 
\sqrt{n_c} &-& i\hbar^2{\sqrt{n_c}\over 
2m}\nabla^2\theta+{\hbar^2\sqrt{n_c}\over 
2m}(\mbox{\boldmath$\nabla$}\theta)^2 \nonumber \\
&-&{i\hbar^2\over m}(\mbox{\boldmath$\nabla$}\sqrt{n_c})\cdot 
\mbox{\boldmath$\nabla$}\theta,\label{eq:dynamics16}
\eea
where we have defined
\begin{equation}
\mu_c({\bf r}, t)\equiv-{\hbar^2\nabla^2\sqrt{n_c}\over 2m\sqrt{n_c}} 
+ V_{ex}({\bf r}) + gn_c (\br, t).
\label{eq:dynamics17}
\end{equation}
Separating out the real and imaginary path gives two equations.  The 
real part gives
\begin{equation}
\hbar{\partial\theta\over\partial t}=-(\mu_c+{1\over 2} 
m\bv_c^2),\label{eq:dynamics18}\end{equation}
where the condensate velocity field is defined by
\begin{equation}m{\bf v}_c ({\bf r}, t) 
\equiv\hbar\mbox{\boldmath$\nabla$}\theta ({\bf r}, t).
\label{eq:dynamics19}\end{equation}
The imaginary part of (\ref{eq:dynamics16}) gives
\begin{equation}
i{\partial\sqrt{n_c}\over\partial t} = - {\sqrt{n_c}\over 
2}\mbox{\boldmath$\nabla$}\cdot{\bf v}_c 
-{\bf v}_c \mbox{\boldmath$\nabla$}\sqrt{n_c}, 
\label{eq:dynamics20}\end{equation}
which can be rewritten in the form
\begin{equation}
{\partial n_c\over\partial t}=-n_c\mbox{\boldmath$\nabla$}
\cdot{\bf v}_c -{\bf v}_c 
\cdot\mbox{\boldmath$\nabla$}n_c=-\mbox{\boldmath$\nabla$}
\cdot (n_c{\bf v}_c). \label{eq:dynamics21}
\end{equation}
>From now on, we shall set $\hbar =1$ except in some final formulas.  

In summary, we have shown that the time-dependent GP equation for 
the  two-component order parameter $\Phi({\bf r}, t)$ is completely 
equivalent to the following two coupled equations for the condensate 
density $n_c({\bf r}, t)$ and velocity field $\bv_c({\bf r}, t)$:
\bea
&&{\partial n_c\over\partial 
t}=-\mbox{\boldmath$\nabla$}\cdot(n_c{\bf v}_c)\nonumber\\
&&m\left({\partial{\bf v}_c\over\partial t} +{1\over 2}  
\mbox{\boldmath$\nabla$}{\bf v}_c^2\right) = - 
\mbox{\boldmath$\nabla$}\mu_c\ ,\label{eq:dynamics22}
\eea
where the position and time-dependent generalized condensate chemical 
potential $\mu_c$ is defined in (\ref{eq:dynamics17}).  In this 
formulation, a complete description of the condensate dynamics is 
given in terms of two variables $n_c({\bf r}, t)$ and ${\bf v}_c({\bf 
r},t)$, reminiscent of the hydrodynamic equations for a classical 
fluid.  In the recent BEC literature, the equations in 
(\ref{eq:dynamics22}) are often referred to as the ``hydrodynamic'' 
theory.  A better description would be to call it the ``quantum 
hydrodynamic'' theory, since it is equivalent to the mean-field GP 
equation.
In later sections, we shall see that an exact GP equation taking into 
account the dynamics of the non-condensate leads to a generalized set 
of equations quite analogous to (\ref{eq:dynamics22}).  The key 
equation (\ref{eq:dynamics19}) defining the superfluid velocity field 
of the condensate will turn out to be quite general.  Needless to 
say, all aspects related to superfluidity of a trapped Bose gas (as 
compared to Bose condensation) are tied to the fact that the 
condensate exhibits motion related to the gradient of a phase (which 
means that, ignoring vortices, the condensate motion is irrotational, 
$\mbox{\boldmath$\nabla$}\times{\bf v}_c({\bf r}, t)=0).$

Stringari~\cite{Sstringari} first pointed out that within the {\it 
dynamic} TF approximation, one could combine the two equations in 
(\ref{eq:dynamics22}) into a single condensate wave equation.  This 
approach will be the basis of our development in these lectures.  
Neglecting the kinetic energy term $-\nabla^2\sqrt{n_c}$ in
(\ref{eq:dynamics17}) as small compared to the condensate interaction 
energy $gn_c$, we linearize the resulting equations around the 
equilibrium values
\bea n_c &=& n_{c0} +\delta n_c\nonumber\\
{\bf v}_c &=& {\bf v}_{c0} +\delta{\bf v}_c\ .
\label{eq:dynamics23}
\eea
We obtain
\bea
{\partial\delta n_c\over\partial t}&=& - 
\mbox{\boldmath$\nabla$}\cdot(n_{c0}\delta{\bf 
v}_c)-\mbox{\boldmath$\nabla$}\cdot({\bf v}_{c0}\delta n_c)\nonumber\\
{\partial\delta {\bf v}_c\over\partial t} &=& - 
\mbox{\boldmath$\nabla$}\left(\mu^\prime_{c0} +g\delta n_c + m{\bf 
v}_{c0}\cdot\delta{\bf v}_c\right)\, ,\label{eq:dynamics24}
\eea
where $\mu^\prime_{c0} \equiv\mu_{c0}+{1\over 2}m{\bf v}^2_{c0}$ is 
independent of position.  Assuming that there is no vortex in the 
solution of the static GP equation (ie, ${\bf v}_{c0}=0),$ 
(\ref{eq:dynamics24}) reduces to the following coupled linearized 
equations
for $\delta n_c$ and $\delta{\bf v}_c$:
\bea
{\partial\delta n_c\over\partial t} &=& 
-\mbox{\boldmath$\nabla$}\cdot(n_{c0}({\bf r})\delta{\bf 
v}_c)\nonumber\\
{\partial\delta{\bf v}_c\over \partial t} &=& -{g\over m} 
\mbox{\boldmath$\nabla$}\delta n_c .
\label{eq:dynamics25}
\eea
These can be combined to give the well known $T=0$ Stringari wave 
equation~\cite{Sstringari,Dalgiorpit}
\begin{equation}
{\partial^2\delta n_c\over\partial t^2} = {g\over 
m}\mbox{\boldmath$\nabla$}\cdot\left[n_{c0}({\bf 
r})\mbox{\boldmath$\nabla$}\delta n_c\right].\label{eq:dynamics26}
\end{equation}

Since the derivation used the TF approximation (where $n_{c0}({\bf 
r})$ is given by (\ref{eq:dynamics11}) for $r\le R_{TF}),$ one can 
rewrite (\ref{eq:dynamics26}) in the equivalent form (for an 
isotropic trap potential)
\begin{figure}
  \centerline{\epsfig{file=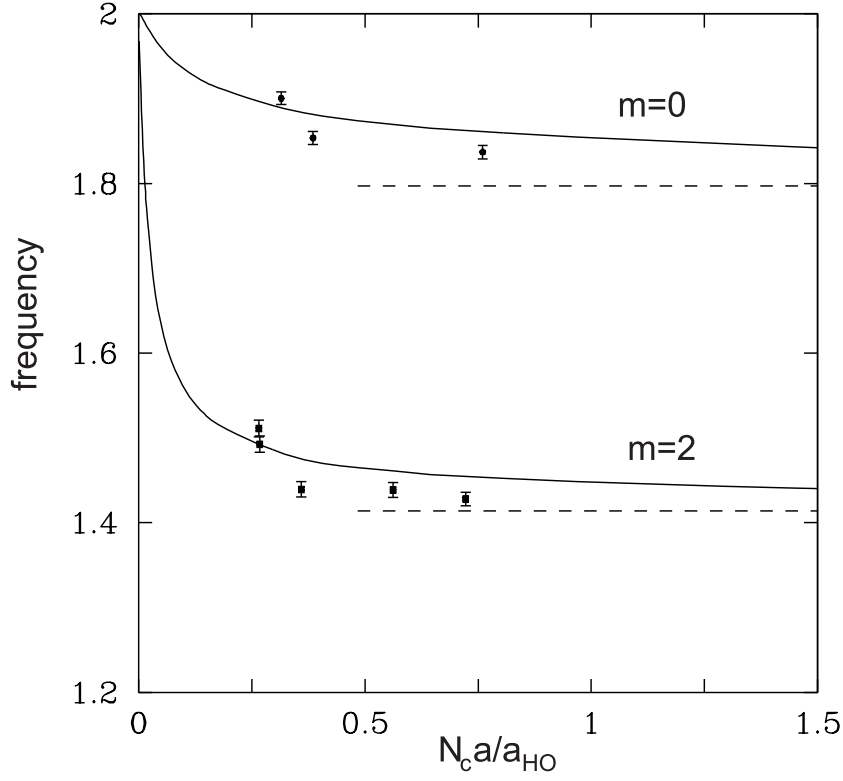,width=4.4in}}
\caption{Plot of the lowest condensate mode frequencies (monopole and 
quadrupole) as a
function of $N_c$. The full curves are the solutions of the coupled
Bogoliubov equations$^{14}$ and the dashed lines are the Stringari TF
approximation Ref.~\protect\cite{Sstringari}. The data points are for 
a JILA trap with $\lambda=\sqrt 8 $~\protect\cite{cornell}. For 
these trap parameters, $N_c$ varies from approximately $10^3$ to
$10^4$ atoms. For further details, see Fig.14 of 
Ref.~\protect\cite{Dalgiorpit}
..} \label{fig:canberra1}
\end{figure}

\begin{equation}
{\partial^2\delta n_c\over\partial t^2} = {\mu_{c0}\over 
m}\mbox{\boldmath$\nabla$}\cdot\left\{\left[
[1-{r^2\over R^2_{TF}}\right]\mbox{\boldmath$\nabla$}\delta 
n_c\right\},\ \ r\le R_{TF},\label{eq:dynamics27}
\end{equation}
where $\mu_{c0}$ is given by (\ref{eq:dynamics13}).  It turns out 
that the normal mode solutions $\delta n_c({\bf r}, t) = \delta 
n_\omega({\bf r})e^{-i\omega t}$ of (\ref{eq:dynamics27})
have frequencies which are independent of the interaction strength 
$g$ or the value of $N_c$.  This is a feature of the underlying TF 
approximation, which typically starts to breakdown (as noted earlier) 
when $N_c\,\lapx\, 10^4$ atoms.  This is shown by explicit numerical 
solutions~\cite{Hutzargrif,Edrupbur} of the coupled Bogoliubov 
equations of motion which describe the normal mode solutions of the 
linearized GP equation when we take the kinetic energy of the 
condensate amplitude $\sqrt{n_c}$ fully into account.  For values of 
$N_c\, \lapx\, 10^4$, the normal mode frequencies depend 
significantly on the magnitude of $N_c$, as shown in 
Fig.~\ref{fig:canberra1}.

We also note that (\ref{eq:dynamics26}) can be equally well rewritten 
in terms of the superfluid velocity $\delta{\bf v}_c $ or, 
equivalently, the phase fluctuations $\delta\theta$. This emphasizes 
that the measured condensate density 
fluctuations~\cite{cornell,ketterle} are directly related to the 
existence of phase fluctuations.  The existence of collective modes 
of a pure condensate may thus be viewed already as ``evidence'' of 
superfluidity, the latter being always a consequence of the phase 
coherence of the macroscopic wavefunction given in 
(\ref{eq:dynamics15}) which gives rise to the irrotational velocity 
in (\ref{eq:dynamics19}).  

We conclude this section with several 
examples~\cite{Dalgiorpit,Sstringari,Alfet} of condensate normal 
modes based on solving the $T=0$ Stringari equation 
(\ref{eq:dynamics26}). A wonderful aspect about the collective 
oscillations of a condensate in a trapped gas is you can ``see'' 
them.   As Ketterle has remarked, these condensates are {\em robust} 
- one can kick them, shake them and these ``wisps'' of Bose-condensed 
matter keep their integrity.~\cite{cornell,ketterle}
 The uniform Bose-condensed gas is especially simple, since 
$\delta n_c = \delta n_{k\omega}e^{i({\bf k}
\cdot{\bf r}-\omega t)}.$  This gives
\begin{equation}
-\omega^2\delta n_{k\omega} ={gn_{c0}\over m}(-k^2)\delta 
n_{k\omega},\label{eq:dynamics28}
\end{equation}
or $\omega^2 = c^2_0k^2,$ with $c_0\equiv\sqrt{gn_{c0}/m}.$  This 
recovers the well-known Bogoliubov phonon oscillations~\cite{Nnbog} 
of a uniform Bose condensate.  The neglect of the kinetic energy in 
our TF approximation precludes us from obtaining the particle-like 
behaviour at large values of the wavevector $k$.~\cite{Alfet}

The Kohn (or sloshing) mode corresponds to the oscillation of the 
centre-of-mass of the static condensate profile with the trap 
frequency $\omega_0.$  This mode is described by
\begin{equation}
n_c({\bf r}, t) = n_{c0}({\bf r}- \mbox{\boldmath$\eta$}(t)), 
\label{eq:dynamics29}
\end{equation}
where 
$\frac{d \mbox{\boldmath$\eta$}(t)}{dt} \equiv{\bf v}_c(t)$ and
\beq {d^2 \mbox{\boldmath$\eta$}(t)\over dt^2}= -\omega^2_0 
\mbox{\boldmath$\eta$}(t).\label{eq:dynamics30}
\eeq
The proof is simple.  Linearizing (\ref{eq:dynamics29}), we find
\begin{equation}
n_c({\bf r}, t) = n_{c0}({\bf r}) - 
\mbox{\boldmath$\eta$}\cdot\mbox{\boldmath$\nabla$} n_{c0}({\bf 
r}),\label{eq:dynamics31}
\end{equation}
which gives the explicit form for the condensate fluctuation,
\begin{equation}
\delta n_c({\bf r},t) = {1\over g}m\omega^2_0{\bf 
r}\cdot\mbox{\boldmath$\eta$}(t).\label{eq:dynamics32}
\end{equation}
Inserting this into the Stringari equation (\ref{eq:dynamics26}), we 
obtain
\bea
-\mbox{\boldmath$\nabla$}n_{c0}\cdot 
{d^2\mbox{\boldmath$\eta$}(t)\over dt^2} &=& 
\omega^2_0\mbox{\boldmath$\nabla$}\cdot 
\left[n_{c0}\mbox{\boldmath$\nabla$}(\mbox{\boldmath$\eta$} 
(t)\cdot{\bf r})\right]\nonumber\\
&=&\omega^2_0\mbox{\boldmath$\nabla$}\cdot\left[n_{c0}
\mbox{\boldmath$\eta$}\right]\nonumber\\
&=&\omega^2_0\mbox{\boldmath$\eta$}\cdot
\mbox{\boldmath$\nabla$}n_{c0}.\label{eq:dynamics33}
\eea
This confirms that the centre-of-mass position 
$\mbox{\boldmath$\eta$}(t)$ of the static condensate distribution 
satisfies the SHO equation (\ref{eq:dynamics30}) with frequency 
$\omega_0$.

The breathing (or monopole) condensate normal mode corresponds to a 
velocity fluctuation of the form
\begin{equation}
\delta{\bf v}_c({\bf r}, t) = A{\bf r} \,e^{-i\omega t}\ (r \le 
R_{TF}).\label{eq:dynamics34}
\end{equation}
Using $\mbox{\boldmath$\nabla$}\cdot{\bf r} = 3$ and 
$\mbox{\boldmath$\nabla$}\cdot(r^2{\bf r})=5r^2,$ it is easy to 
verify from the continuity equation in (\ref{eq:dynamics22}) that the 
associated density fluctuation is
\bea
-i\omega\delta n_c&=&-\mbox{\boldmath$\nabla$}\cdot
\left[n_{c0}({\bf r})\delta{\bf v}_c\right]\nonumber\\
&=&-{\mu_{c0}\over g}A\left[3-5{r^2\over R^2_{TF}}\right]
e^{-i\omega t},\label{eq:dynamics35}
\eea
or
\begin{equation}
\delta n_{\omega}({\bf r}) \equiv B\left(1-{5\over 3} {r^2\over 
R^2_{TF}}\right),\label{eq:dynamics36}
\end{equation}
where $\mu_{c0}$ is given by (\ref{eq:dynamics12}).  Inserting this 
into the Stringari wave equation (\ref{eq:dynamics27}) gives
\bea
-\omega^2\delta n_\omega({\bf r}) &=&{\mu_{c0}\over 
m}\mbox{\boldmath$\nabla$}\cdot\left[\left(1-{r^2\over 
R^2_{TF}}\right)B{5\over 3}{2{\bf r}\over R^2_{TF}}\right]\nonumber\\
&=&-5\omega^2_0\delta n_\omega({\bf r}).\label{eq:dynamics37}
\eea
Thus the breathing mode for an isotropic parabolic trap has a 
frequency $\omega =\sqrt 5\, \omega_0.$  As noted earlier, this mode 
frequency is not explicitly dependent on the interaction strength 
$g$.  However the underlying theory is very dependent on mean-field 
effects.  Moreover, we recall that a {\it non-interacting} trapped 
Bose gas has a breathing mode with a frequency $\omega=2\omega_0$ at 
all temperatures.~\cite{Dalgiorpit}

As a final example, we consider the so-called ``surface'' modes of a 
$T=0$ condensate, as described by~\cite{Sstringari}\begin{equation}
\delta {\bf v}_c({\bf r}, t) = A\mbox{\boldmath$\nabla$}\left[r^l 
Y_{lm}(\theta, \phi)\right] e^{-i\omega t}.\label{eq:dynamics38}
\end{equation}
This mode corresponds to phase fluctuations of the form 
$\delta\theta_\omega =mA\,r^l Y_{lm}(\theta, \phi)$.  One may easily 
verify that  $\mbox{\boldmath$\nabla$}\cdot\delta{\bf v}_c=0$ and 
hence from (\ref{eq:dynamics25}) we obtain
\begin{equation}
{\partial^2\delta{\bf v}_c\over\partial t^2} = 
-\omega^2_0\mbox{\boldmath$\nabla$}(\delta{\bf v}_c\cdot{\bf 
r}).\label{eq:dynamics39}
\end{equation}
This is equivalent to
\begin{equation}
{\partial^2\delta\theta\over\partial t^2} 
=-\omega^2_0(\mbox{\boldmath$\nabla$}\delta\theta)\cdot{\bf 
r}\label{eq:dynamics40}
\end{equation}
or
\begin{equation}
-\omega^2\delta\theta_\omega({\bf r}) =-\omega^2_0 
l\delta\theta_\omega(\br).\label{eq:dynamics41}
\end{equation}
Thus the surface oscillations of the condensate phase have a 
frequency given by $\omega=\sqrt{l}\omega_0 (l=1,2,3,\dots).$

One great advantage of the quantum hydrodynamic formalism (within the 
TF approximation) is that it is easy to also treat the $T=0$ normal 
modes of an anisotropic parabolic well.  We refer to the literature 
for further discussion.~\cite{Dalgiorpit,Sstringari,Alfet}  We also 
note that it is straightforward to derive a Stringari wave equation 
for the oscillations of a vortex state (this has been done by 
Svidzinsky and Fetter~\cite{ALfett}).

\section{Generalized GP equation at finite temperatures}
\label{sec:temperatures}

In this section, we generalize the $T=0$ GP equation of 
Section~\ref{sec:Pure} to finite $T$, so that it includes the effect 
of the coupling of the condensate to the non-condensate degrees of 
freedom.  We concentrate in this Section on how the $T=0$ condensate 
modes are renormalized and damped by the thermal cloud.  We defer 
discussion of the appearance of new collective modes mainly 
associated with  the thermal cloud to sections~\ref{sec:coupled} and 
\ref{sec:fluid}.

Up to the beginning of 2000, the study of the dynamics of a trapped 
Bose gas at finite temperatures has been largely ignored by 
experimentalists but actively studied by many theorists (especially 
in the last year or so).  One of the reasons for the lack of finite 
temperature data is that there are so many interesting phenomena to 
study at $T\simeq 0$!  However, another reason seems to be the 
implicit belief that the presence of the non-condensate just 
complicates the behaviour of a pure $T=0$ condensate - but is not the 
source of any interesting new physics.  In this and subsequent 
sections, I hope to argue that this ``belief''  is quite wrong.  The 
coupling of the condensate and non-condensate degrees of freedom at 
finite $T$ leads to a new two-component system in which {\em both} 
components can exhibit coherent behaviour, analogous to the 
well-known superfluid macroscopic phenomena appearing in liquid 
$^4$He.

We shall see that, as expected, the finite temperature GP equation of 
motion for $\Phi({\bf r}, t)$ is not closed.  It's general solution 
involves knowing the equations of motion for the non-condensate.  We 
will work within an approximation where the non-condensate atoms can 
be described by a quantum Boltzmann transport equation for the 
single-particle distribution function $f({\bf p}, {\bf r}, t),$ with 
\begin{equation}
{\tilde n}({\bf r}, t)=\int{d{\bf p}\over (2\pi)^3} f({\bf p}, {\bf 
r}, t).\label{eq:temperatures42}\end{equation}
The details of this kinetic equation will be developed in 
Section~\ref{sec:coupled}.  In this section, we concentrate on 
incorporating the effect of a {\em static} thermal cloud into the 
finite $T$ dynamics of the condensate.

The theory of interacting Bose-condensed fluids is most usefully 
discussed using quantum field operators.  This procedure was 
formalized by Beliaev~\cite{Bel2} in 1957 and developed by 
Bogoliubov~\cite{Bogg}, Gavoret and Nozi\`{e}res~\cite{Gavnoz}, 
Martin and Hohenberg~\cite{Hohmar}, and others in the early 1960's.  
We recall:
\begin{eqnarray}
& {\hat\psi}^\dagger({\bf r})= \mbox{creates atom at} \ {\bf r} 
\nonumber \\
&{\hat\psi}({\bf r}) =\mbox{destroys atom at} \ {\bf 
r}.\label{eq:temperatures43}
\end{eqnarray}
These quantum field operators satisfy the usual Bose commutation 
relations, such as
\begin{equation}
\left[{\hat\psi}({\bf r}), {\hat\psi}^\dagger({\bf r}^\prime)\right] 
= \delta({\bf r} - {\bf r}^\prime). \label{eq:temperatures44}
\end{equation}
All observables can be written in terms of these quantum field 
operators, such as the density ${\hat n}(\br) 
={\hat\psi}^\dagger(\br){\hat\psi}(\br)$ and interaction energy
\begin{eqnarray}
{\hat V} &=& {1\over 2}\int d{\bf r} \int d{\bf r}^\prime 
{\hat\psi}^\dagger ({\bf r}^\prime){\hat\psi}^\dagger({\bf r}) v({\bf 
r}-{\bf r}^\prime){\hat\psi}({\bf r}^\prime){\hat\psi}({\bf 
r})\nonumber \\
&=& {1\over 2}g \int d{\bf r} {\hat\psi}^\dagger({\bf 
r}){\hat\psi}^\dagger({\bf r}){\hat\psi}({\bf r}){\hat\psi}({\bf 
r}).\label{eq:temperatures45}
\end{eqnarray}

The crucial idea due to Bogoliubov~\cite{Nnbog,Bogg} and later 
generalized by Beliaev~\cite{Bel2} is to separate out the condensate 
component of the field operators,\begin{equation}
{\hat\psi}({\bf r}) = \left<{\hat\psi}({\bf r})\right> 
+{\tilde\psi}({\bf r}), \label{eq:temperatures46}
\end{equation}
where
\begin{displaymath}
\left<{\hat\psi}({\bf r}) \right> \equiv \Phi({\bf r}) = \mbox{Bose 
macroscopic wavefunction}.\nonumber
\end{displaymath}
This quantity plays the role of the ``order parameter'' for the Bose 
superfluid phase transition:
\bea
\Phi({\bf r}) &= 0 \ \ \ \ T > T_c \nonumber \\
&\ne 0 \ \ \ \ T < T_c. \nonumber\eea

We note that $\Phi({\bf r}) \equiv\sqrt{n_c} e^{i\theta}$ is a 
2-component order parameter.  Clearly, $\Phi({\bf r})$ is not simply 
related to the many-particle wavefunctions $\Psi({\bf r}_1, {\bf 
r}_2, \dots {\bf r}_N).$
The thermal average in $\left<{\hat\psi}({\bf r})\right>$ involves a 
small symmetry-breaking perturbation to allow $\Phi(\br)$ to be 
finite,
\begin{equation}
{\hat H}_{SB} = \lim_{\eta\to 0}\int d{\bf r}\left[\eta ({\bf 
r}){\hat\psi}^\dagger({\bf r}) + \eta^\ast({\bf r}) {\hat\psi}({\bf 
r})\right].\label{eq:temperatures47}
\end{equation}
The philosophy behind the concept of symmetry-breaking was 
extensively discussed by Bogoliubov~\cite{Bogg}, for a variety of 
condensed matter systems, in an article which is still highly 
recommended.

It is useful to make a few comments on the physics behind $\Phi({\bf 
r}, t)$.  $\Phi ({\bf r}, t)$ is a {\it coherent} state, with a 
``clamped'' value of phase - rather than a Fock-state of fixed $N$, 
with no well-defined phase.  $\Phi({\bf r}, t)$ acts like a {\it 
classical} field, since quantum fluctuations are negligible when 
$N_c$ is large.  Probably P.W. Anderson~\cite{PWand} deserves the 
greatest credit for understanding (in the period 1958-1963) the new 
physics behind working with a broken-symmetry state $\Phi({\bf r}, 
t)$, both in BCS superconductors and in superfluid $^4$He.   It 
captures the physics of the new phase of matter (such as the 
occurrence of the Josephson effect) and the associated superfluidity.  
The symmetry-breaking perturbation~(\ref{eq:temperatures47}) allows 
the system to internally set up off-diagonal symmetry-breaking 
fields, which persist even when the external symmetry-breaking 
perturbation in (\ref{eq:temperatures47}) is set to zero at the end 
$(\eta \rightarrow 0)$.  The same sort of physics is the basis of  
the BCS theory of superconductors.

One can formulate the GP and Bogoliubov approximations directly in 
terms of a variational many-particle wavefunction (see the lectures 
by Leggett in this volume).  However, such formulations are limited 
to simple mean-field approximations.  The explicit introduction of 
the broken-symmetry order parameter $\Phi(\br, t)$ gives a systematic 
way~\cite{Bel2,Bogg,Hohmar,Gorkov} of isolating the role of the Bose 
condensate in a general treatment of an interacting Bose-condensed 
fluid.  As we shall see, the resulting formalism allows one to deal 
with questions related to damping as well as superfluidity in both 
the collisionless and hydrodynamic regions.

The exact Heisenberg equation of motion for the field operator is
\begin{eqnarray}
i{\partial{\hat\psi}({\bf r}, t)\over\partial t} &=& 
\left[-{\nabla^2\over 2m}+V_{ex}({\bf r}) + \delta U({\bf r}, 
t)\right]
{\hat\psi}({\bf r}, t)\nonumber \\
&+&\eta({\bf r})+g{\hat\psi}^\dagger ({\bf r}, t){\hat\psi}({\bf r}, 
t){\hat\psi}({\bf r}, t),\label{eq:temperatures48}
\end{eqnarray}
where $\delta U({\bf r}, t)$ is a small time-dependent driving 
potential.  This gives an {\em exact} equation of motion for 
$\Phi({\bf r}, t) \equiv\left<{\hat\psi}({\bf r}, t)\right>,$
\begin{eqnarray}
i{\partial\Phi({\bf r}, t)\over\partial t} &=&\left[-{\nabla^2\over 
2m}+V_{ex}({\bf r})
+\delta U({\bf r}, t)\right]\Phi({\bf r}, t)\nonumber \\
&+&\eta({\bf r})+g\left<{\hat\psi}^\dagger ({\bf r}, 
t){\hat\psi}({\bf r}, t){\hat\psi}({\bf r}, 
t)\right>,\label{eq:temperatures49}
\end{eqnarray}
with [(using the decomposition (\ref{eq:temperatures46})]
\begin{equation}
{\hat\psi}^\dagger{\hat\psi}{\hat\psi} = |\Phi|^2\Phi + 
2|\Phi|^2{\tilde\psi}+\Phi^2{\tilde\psi}^\dagger + 
\Phi^\ast{\tilde\psi}{\tilde\psi}+2\Phi{\tilde\psi}^\dagger{\tilde\psi} 
+ {\tilde\psi}^\dagger{\tilde\psi}{\tilde\psi}.
\label{eq:temperatures50}
\end{equation}

Taking the symmetry-breaking average of (\ref{eq:temperatures50}), 
one finds
\begin{equation}
\left<{\hat\psi}^\dagger{\hat\psi}{\hat\psi} \right> = 
n_c\Phi+{\tilde m}\Phi^\ast+2{\tilde 
n}\Phi+\left<{\tilde\psi}^\dagger  
{\tilde\psi}{\tilde\psi}\right>,\label{eq:temperatures51}
\end{equation}
where
\bea
n_c ({\bf r}, t) &\equiv& |\Phi({\bf r}, t)|^2 = \mbox{condensate 
density} \nonumber\\
{\tilde n}({\bf r}, t) &\equiv&  \left<{\tilde\psi}^\dagger({\bf r}, 
t){\tilde\psi}({\bf r}, t)
\right> = \mbox{non-condensate density} \nonumber\\
{\tilde m}({\bf r}, t) &\equiv&\left<{\tilde\psi}({\bf r}, 
t){\tilde\psi}({\bf r}, t)
\right> = \mbox{off-diagonal (anomalous) 
density}.\label{eq:temperatures52}
\eea
Using (\ref{eq:temperatures51}) in (\ref{eq:temperatures49}), our 
``exact'' equation of motion for $\Phi({\bf r}, t)$ 
is~\cite{Zarnikgrif}
\bea i{\partial\Phi\over\partial t} &=& \left[-{\nabla^2\over 
2m}+V_{ex} +gn_c({\bf r}, t) + 2g{\tilde n}({\bf r}, 
t)\right]\Phi\nonumber\\
&+& g{\tilde m}({\bf r}, t)\Phi^\ast 
+g\left<{\tilde\psi}^\dagger({\bf r},t){\tilde\psi}({\bf 
r},t){\tilde\psi}({\bf r},t)\right>.\label{eq:temperatures53}
\eea
We now consider various approximations to the generalized GP equation 
in (\ref{eq:temperatures53}):
\begin{enumerate}\item[(a)] The Hartree-Fock-Bogoliubov (HFB) 
approximation for $\Phi$ corresponds to neglecting the three-field 
correlation function 
$\left<{\tilde\psi}^\dagger{\tilde\psi}\psi\right>$ but keeping the 
$n_c, {\tilde n}$ and ${\tilde m}$ fluctuations.  This HFB has been 
exhaustively treated in many recent papers, in conjunction with the 
separate equations of motion for ${\tilde n}$ and ${\tilde 
m}$.~\cite{Griffin,Tomgrif,Sgior}  This HFB approximation can be used 
to generate the same normal mode spectrum which the Beliaev 
second-order self-energy approximation gives at finite 
$T$.~\cite{Huashigrif}
\item[(b)] The {\em dynamic} Popov approximation corresponds to 
ignoring both 
$\left<{\tilde\psi}^\dagger{\tilde\psi}{\tilde\psi}\right>$ and 
${\tilde m}=\left<{\tilde\psi}{\tilde\psi}\right>$ in 
(\ref{eq:temperatures53}).  Theories of this kind involve coupled 
equations for $\Phi$ and ${\tilde n}$.~\cite{Mingtosi}
\item[(c)] The {\em static} Popov 
approximation~\cite{Griffin,Hutzargrif,Dalgiorpit} involves even a 
further simplification, namely it ignores fluctuations in the density 
${\tilde n}({\bf r}, t)$ of the thermal cloud,
\begin{equation}
{\tilde n}({\bf r}, t)\simeq n_{c0}({\bf 
r}).\label{eq:temperatures54}\end{equation}
\end{enumerate}
Using (\ref{eq:temperatures54}) in (\ref{eq:temperatures53}) 
corresponds to treating the dynamics of the condensate moving in the 
{\em static} mean field of the non-condensate thermal cloud, ie,
\begin{equation}
i\hbar{\partial\Phi\over\partial t}= \left[-{\nabla^2\over 2m} 
+V_{ex}({\bf r})+2g{\tilde n}_0({\bf r})+gn_c({\bf r}, 
t)\right]\Phi({\bf r}, t).\label{eq:temperatures55}\end{equation}
There is a considerable literature based on this static 
Popov-approximation.~\cite{Dalgiorpit}  In a nutshell, in this 
section, we will discuss an extension of (\ref{eq:temperatures55}) 
which involves including the damping associated with the 
$\langle{\tilde\psi}^\dagger{\tilde\psi}{\tilde\psi}\rangle$ term, 
but again treating the thermal cloud statically.

As in Section~\ref{sec:Pure}, we first rewrite 
(\ref{eq:temperatures53}) in terms of the condensate amplitude and 
phase.  Inserting (\ref{eq:dynamics15}) into 
(\ref{eq:temperatures53}) and following the same procedure which led 
to the $T=0$ equations in (\ref{eq:dynamics22}), we obtain:

\begin{equation}
{\partial n_c\over\partial t}+\mbox{\boldmath$\nabla$}\cdot n_c{\bf 
v}_c = 2gIm\left[\Phi^{\ast 2}{\tilde m}+\Phi^\ast 
\langle{\tilde\psi}^\dagger{\tilde\psi}{\tilde\psi}
\rangle\right]\label{eq:temperatures56}
\end{equation}

\begin{equation}
{\partial\theta\over\partial t} = -\left(\mu_c + 
{(\mbox{\boldmath$\nabla$}\theta)^2\over 
2m}\right),\label{eq:temperatures57}
\end{equation}
where now (compare with (\ref{eq:dynamics17}))

\bea
\mu_c({\bf r}, t) = &-& {\nabla^2\sqrt{n_c}\over 
2m\sqrt{n_c}}+V_{ex}+gn_c +2g{\tilde n}\nonumber\\
&+& {g\over n_c} Re\left[\Phi^{\ast 2}{\tilde 
m}+\Phi^\ast\left<{\tilde\psi}^\dagger{\tilde\psi}
{\tilde\psi}\right>\right].
\label{eq:temperatures58}
\eea
Formally, (\ref{eq:temperatures53}) and hence 
(\ref{eq:temperatures56}) - (\ref{eq:temperatures58}) are {\it exact} 
results.  In these lectures, we will limit ourselves to finite 
temperatures where the dominant thermal excitations are well 
approximated by high energy non-condensate atoms moving in a 
self-consistent dynamic HF mean-field
\bea
{\tilde\varepsilon}_p({\bf r}, t) &=& {p^2\over 2m} + V_{ex}({\bf 
r})+2g\left[n_c({\bf r}, t)+{\tilde n}({\bf r}, t)\right]\nonumber\\
&\equiv& {p^2\over 2m} + U({\bf r}, t).\label{eq:temperatures59}\eea
In this region, we shall neglect the anomalous pair correlations 
among the thermal atoms (ie, ${\tilde m} = 0$).  One can show that 
both ${\tilde m}$ and
$\langle{\tilde\psi}^\dagger{\tilde\psi}{\tilde\psi}\rangle$
in (\ref{eq:temperatures56}) and (\ref{eq:temperatures58}) are of 
order $g$.  Thus one sees the correction terms in 
(\ref{eq:temperatures56}) and (\ref{eq:temperatures58}) involving 
these functions are of order $g^2$.  In fact, calculation shows that 
to order $g$, 
$\langle{\tilde\psi}^\dagger{\tilde\psi}{\tilde\psi}\rangle$ is 
imaginary.  Our procedure is to ignore the $O(g^2)$ terms in 
(\ref{eq:temperatures56}) and (\ref{eq:temperatures58}) except those 
which are imaginary and hence are a source of damping of condensate 
motion.

The end result of this approach is that we are left with
\bea
{\partial n_c\over\partial t}+\mbox{\boldmath$\nabla$}\cdot n_c{\bf 
v}_c &=&-\Gamma_{12}[f, \Phi]\nonumber\\
m\left({\partial{\bf v}_c\over\partial t} + {1\over 
2}\mbox{\boldmath$\nabla$}{\bf v}_c^2\right) &=& 
-\mbox{\boldmath$\nabla$}\mu_c\ ,\label{eq:temperatures60}
\eea
with
\begin{equation}
\mu_c({\bf r}, t) = -{\nabla^2\sqrt{n_c}\over 2m\sqrt{n_c}} + 
V_{ex}({\bf r}) + gn_c({\bf r}, t) + 2g{\tilde n}({\bf r}, 
t)\label{eq:temperatures61}
\end{equation}
and we have defined the new function
\begin{equation}
\Gamma_{12}[f, \Phi] \equiv - 2g 
Im\left[\Phi^\ast\langle{\tilde\psi}^\dagger{\tilde\psi}
{\tilde\psi}\rangle\right].
\label{eq:temperatures62}
\end{equation}

We see that $\Gamma_{12}$ plays the role of a source term in the 
condensate continuity equation in (\ref{eq:temperatures60}) .  It 
depends on both $\Phi({\bf r}, t)$ and the single-particle 
distribution function $f({\bf p}, {\bf r}, t).$  It is useful to note 
that (\ref{eq:temperatures60}) and (\ref{eq:temperatures61}) are 
equivalent to the approximate GP equation (see also 
Refs.~\cite{Stoof99,Walser99})
\begin{equation}
i\hbar{\partial\Phi\over\partial t} = \left[-{\hbar^2\nabla^2\over 
2m}+V_{ex} ({\bf r}) + gn_c({\bf r}, t) + 2g{\tilde n}({\bf r}, t) 
-i\hbar R({\bf r}, t)\right]\Phi\ 
,\label{eq:temperatures63}\end{equation}
where the dissipative function is
\begin{equation}
R({\bf r}, t) \equiv{\Gamma_{12}[f, \Phi]\over 2n_c({\bf r}, t)} \sim 
0(g^2).\label{eq:temperatures64}\end{equation}

To proceed, we need to calculate the field correlation 
$\langle{\tilde\psi}^\dagger{\tilde\psi}{\tilde\psi}\rangle$ which 
determines $\Gamma_{12}$ in (\ref{eq:temperatures62}).  This has been 
done in Appendix A of Ref.~\cite{Zarnikgrif}, working to lowest order 
in $g$ at finite $T$ where the single-particle spectrum in 
(\ref{eq:temperatures59}) is adequate.  The result is
\bea
\langle{\tilde\psi}^\dagger{\tilde\psi}{\tilde\psi}\rangle &=& 
-ig{\Phi({\bf r}, t)\over(2\pi)^5}\int d{\bf p}_1\int d{\bf p}_2\int 
d{\bf p}_3 \nonumber \\
&\times&\delta\left({\bf p}_c+{\bf p}_1-{\bf p}_2-{\bf 
p}_3)\delta(\epsilon_c+{\tilde\epsilon}_1-{\tilde\epsilon}_2
-{\tilde\epsilon}_3\right)\nonumber\\
&\times&\left[f_1(1+f_2)(1+f_3)-(1+f_1)f_2f_3\right],
\label{eq:temperatures65}
\eea
where $f_1\equiv f({\bf p}, {\bf r}, t), \epsilon_c\equiv\mu_c({\bf 
r}, t)+\frac 12 m{\bf v}_c^2({\bf r}, t)$ is the local condensate 
atom energy and ${\bf p}_c\equiv m{\bf v}_c$ is the condensate atom 
momentum.  In Section~\ref{sec:coupled}, we shall see that 
(\ref{eq:temperatures65}) is closely related to the collision 
integral $C_{12}[f,\Phi]$ which enters the Boltzmann equation for 
$f({\bf p}, {\bf r}, t)$ and describes collisions with the 
non-condensate atoms in which one condensate atom is involved.  We 
note that to leading order, the expression in 
(\ref{eq:temperatures65}) is imaginary.  Using this in 
(\ref{eq:temperatures62}), we find
\bea
\Gamma_{12}({\bf r}, t) &=& 2g^2{n_c({\bf r}, t)\over(2\pi)^5}\int 
d{\bf p}_1\int d{\bf p}_2\int d{\bf p}_3 \delta({\bf p}_c+{\bf 
p}_1-{\bf p}_2-{\bf p}_3)\nonumber \\
&\times&\delta\left(\epsilon_c+{\tilde\epsilon}_1-{\tilde\epsilon}
_2-{\tilde\epsilon}_3\right)\nonumber\\
&\times&\left[f_1(1+f_2)(1+f_3)-(1+f_1)f_2f_3\right].
\label{eq:temperatures66}
\eea

As a first application of our new equations in 
(\ref{eq:temperatures60}), we will ignore the dynamics of the 
non-condensate cloud.  That is, we will assume that the condensate 
interacts with a {\it static} thermal cloud in thermal equilibrium

\beq
f(\bp,\br,t) \simeq f^0(\bp,\br) = {1 \over {e^{\beta[p^2 /2 m +
U_0(\br) - \tilde{\mu}_0]}-1}} \,, \label{eq:temperatures67}\eeq 
where ${\tilde\mu}_0$ is the equilibrium chemical potential of the 
non-condensate atoms.  In
Section~\ref{sec:coupled}, we will discuss the $C_{22}$ collisions 
which produce this static equilibrium Bose distribution.  The value 
of ${\tilde\mu}_0$ is known since it must equal the condensate 
equilibrium chemical potential $\mu_{c0}$ given by the static 
solution of the generalized GP equation.  Within the TF 
approximation, (\ref{eq:temperatures61}) reduces to 
\beq \mu_{c0} = V_{\rm ex} (\br)+gn_{c0}(\br) + 2g\tilde{n}_0(\br)
\label{eq:temperatures68}\eeq
and hence we see that (\ref{eq:temperatures59}) reduces to
\bea
{\tilde\epsilon}_p(\br)-{\tilde\mu}_0&=& {p^2\over 2m}+U_0({\bf r}) - 
{\tilde\mu}_0\nonumber \\
&=&{p^2\over 2m}+gn_{c0}({\bf r})\ .\label{eq:temperatures69}\eea
This result is consistent  with the correct Bogoliubov ``excitation 
energy.''~\cite{Dalgiorpit,Alfet} The static non-condensate density 
in (\ref{eq:temperatures68}) is given by
\begin{equation}
{\tilde n}_0({\bf r}) = {1\over\Lambda^3}g_{3/2}\left(z_0=e^{-\beta 
gn_{c0}({\bf r})}\right).
\label{eq:temperatures70}\end{equation}

We can now calculate $\Gamma_{12}({\bf r}, t)$ using the equilibrium 
Bose distribution (\ref{eq:temperatures67}) for the thermal cloud.  
One finds~\cite{Wilgrif} (for a related calculation, see 
Ref.~\cite{Zarnikgrif}).
\beq \Gamma_{12}^0(\br,t) = {n_c(\br,t)
\over \tau_{12}(\br,t)} \big [e^{-\beta (\tilde{\mu}_0 -
\varepsilon_c(\br,t)-{1\over 2}mv_c^2)} - 1 \big ],
\label{eq:temperatures71}
\eeq 
where we have defined a $C_{12}$ collision time
\bea {1 \over
{\tau_{12}(\br,t)}} &\equiv& {2 g^2 \over (2\pi)^5} \!\int
\!\!d{\bf p}_1 \!\int \!\!d{\bf p}_2 \!\int \!\!d{\bf p}_3
\delta(\bp_c+{\bf p}_1-{\bf p}_2-{\bf p}_3) \nonumber \\ &\times&
\delta(\varepsilon_c+\tilde \varepsilon_{p_1} -\tilde
\varepsilon_{p_2}-\tilde \varepsilon_{p_3}) (1+f_1^0) f_2^0 f_3^0 \,.
\label{eq:temperatures72}
\eea
 $\Gamma_{12}[f_0, \Phi]$ in (\ref{eq:temperatures71}) still depends 
on $n_c(\br, t)$ and $\bv_c(\br, t)$ of the condensate through 
$\mu_c, {\bp}_c$ and $\epsilon_c$.
We note that when the condensate is in  static equilibrium, 
$\mu_c({\bf r}, t)\to\mu_{c0}$ and 
${\bf v}_{c0}=0$.  In this case, $\mu_{c0}={\tilde\mu}_0$ and the 
expression in the square bracket in (\ref{eq:temperatures71}) is seen 
to vanish. Thus $\Gamma_{12}(f_0, \Phi_0)=0,$ as it should when both 
components are in static thermal equilibrium.

Summarizing, at this point we have a closed set of equations which 
can be used to describe the dynamics of the condensate in a trapped 
Bose gas which include the interactions with a {\em static} 
equilibrium thermal cloud.  These equations are
\beq
{\partial n_c\over\partial t}+\nabla\cdot n_c{\bf v}_c = 
-\Gamma_{12}[f^0, \Phi]\label{eq:temperatures73}\eeq

\beq
m\left({\partial\over\partial t}+{\bf v}_c\cdot 
\mbox{\boldmath$\nabla$}\right) 
\bv_c=-\mbox{\boldmath$\nabla$}\mu_c\, ,\label{eq:temperatures74}
\eeq
where $\Gamma_{12}[f^0, \Phi]$ is given explicitly by the expression 
in (\ref{eq:temperatures71}) and
\beq
\mu_c(\br, t) = -{\nabla^2\sqrt{n_c}\over 2m\sqrt{n_c}} + 
V_{ex}(\br)  + 2g\tn_0(\br)+gn_c(\br, t).\label{eq:temperatures75}
\eeq
>From (\ref{eq:temperatures73}) and (\ref{eq:temperatures74}), we can 
obtain linearized equations of motion for the
condensate fluctuations $\delta n_c$ and $\delta \bv_c$. We use the
fact that, to lowest order in the fluctuations from static
equilibrium, (\ref{eq:temperatures71}) reduces to 
\beq \delta \Gamma_{12}^0 = {\beta
n_{c0}(\br) \over {\tau_{12}^0(\br)}} \delta \mu_c(\br,t) \, , 
\label{eq:temperatures76}
\eeq
where the ``equilibrium'' $C_{12}$ collision rate [using 
(\ref{eq:temperatures69}) in  (\ref{eq:temperatures72})]
is defined by \bea
{1\over {\tau^0_{12}(\br)}} &\equiv& {2 g^2 \over (2\pi)^5}
\int d{\bf p}_1 \int d{\bf p}_2 \int d{\bf p}_3 \delta({\bf p}_1-{\bf
p}_2-{\bf p}_3) \nonumber \\ &\times&\delta
\big({p_1^2-p_2^2-p_3^2\over{2m}}-g n_{c0}\big ) (1+f_1^0) f_2^0 f_3^0
\,.\label{eq:temperatures77} \eea 
In the static TF approximation, we recall that the equilibrium 
distribution is given by $f_i^0 =
[e^{\beta(p_i^2 /2 m + g n_{c0})}-1]^{-1}$. 

In the present discussion, we further restrict ourselves to the 
dynamic Thomas-Fermi limit valid for large $N_c$, \bea {\partial 
\delta n_c
\over \partial t} +\mbox{\boldmath$\nabla$}\cdot(n_{c0}\delta 
\bv_c)&=& -{1
\over \tau'}\delta n_c
\label{eq:temperatures78} \\ m {\partial \delta \bv_c \over\partial t}
&=&-g\mbox{\boldmath$\nabla$} \delta n_c \,. 
\label{eq:temperatures79} \eea The collision time
$\tau'(\br)$ describes collisions between the condensate and
non-condensate atoms when the condensate is perturbed away from
equilibrium, \beq {1\over\tau'(\br)} = {g n_{c0}(\br) \over
{k_{\rm{B}} T}} {1 \over {\tau^0_{12}(\br)}} \, . 
\label{eq:temperatures80} \eeq The new term on the
right-hand side of (\ref{eq:temperatures78}) causes damping of the 
condensate
fluctuations due to the lack of collisional detailed balance between
the condensate and the static thermal cloud. We note that this
collision time $\tau^\prime(\br) $ is only a function of the position 
$\br$ through its dependence on
the static condensate density $n_{c0}(\br)$. 
\begin{figure}
  \centerline{\epsfig{file=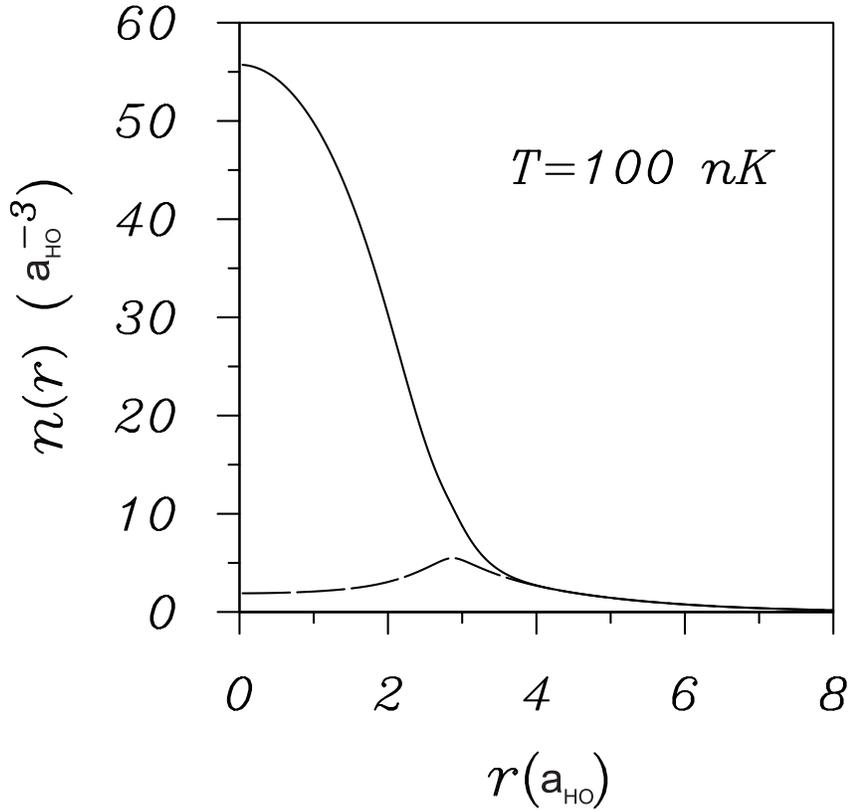,width=4.4in}}
\caption{Density of atoms as a function of position in an isotropic 
parabolic
trap containing 5000 $^{87}$Rb atoms at $T = 100$ nK, with
$T_{\rm{BEC}} = 149$ nK. At this temperature, the total number of 
atoms in the
condensate and non-condensate are equal. The solid line is the total
density and the dashed line is the non-condensate density. The TF 
approximation is not used. The
parameter $a_{\rm{HO}}$ is the oscillator length.  See 
Ref.~\protect\cite{Zarnikgrif}.
} \label{fig:canberra2}
\end{figure}

We can easily combine (\ref{eq:temperatures78}) and 
(\ref{eq:temperatures79}) to obtain what we shall refer
to as the finite $T$ Stringari wave equation~\cite{Wilgrif} \beq 
{\partial^2 \delta
n_c \over \partial t^2} - { g \over m} \mbox{\boldmath$\nabla$}\cdot 
(n_{c0}
\mbox{\boldmath$\nabla$}\delta n_c) = -{1 \over \tau'} {\partial 
\delta n_c
\over \partial t} \,.
\label{eq:temperatures81} \eeq 
 If we
neglect the right-hand side, we obtain the undamped finite $T$
Stringari normal modes $\delta n_c(\br,t) = \delta n_i(\br) e^{-i
\omega_i t}$ given by the solution of \beq - { g
\over m} \mbox{\boldmath$\nabla$}\cdot \big[n_{c0}(\br) 
\mbox{\boldmath$\nabla$}\delta
n_i(\br)\big] = \omega_i^2 \delta n_i(\br).
\label{eq:temperatures82}
\eeq

 As has been noted by several authors in recent
papers~\cite{Dalgiorpit,Sgior}, 
$n_{c0}(\br)$ at finite $T$ can be well approximated by the TF
condensate profile at $T=0$ but with the number of atoms in the
condensate $N_c(T)$ now being a function of temperature.  This is 
because the
static mean field of the non-condensate plays  a minor role. In the 
regions where the condensate density is finite, we effectively always 
have $n_{c0}(\br)\gg{\tilde n}_0(\br),$  as illustrated in 
Fig.~\ref{fig:canberra2}. With
this approximation for $n_{c0}(\br)$, the solutions of the finite $T$
Stringari equation~(\ref{eq:temperatures81}) will be identical to 
those at $T=0$ (see
Section~\ref{sec:Pure}), 
since the $T=0$ Stringari frequencies do not depend on the magnitude
of $N_c$. Of course, as shown by calculations solving the coupled
Bogoliubov equations~\cite{Dalgiorpit,Edrupbur}, when $N_c \,\lapx\,
10^4$ the TF approximation breaks down. Thus the condensate
collective mode frequencies will always become temperature dependent
close to $T_{\rm{BEC}}$, where the TF approximation is no longer
valid.~\cite{Hutzargrif}

We can use the undamped Stringari modes given by 
(\ref{eq:temperatures82}) as a basis set to solve 
(\ref{eq:temperatures81})
and find the damping of these modes. Writing $\delta n_c(\br) = \sum_i
c_i \delta n_i(\br)$, and using the orthonormality condition $\int d
\br \delta n_i(\br) \delta n_j(\br) = \delta_{ij}$, one obtains the
following algebraic equations for the coefficients $c_i$ \beq \omega^2
c_i = \omega_i^2 c_i - i \omega \sum_j \gamma_{ij} c_j \,,
\label{eq:temperatures83}
\eeq
where 
\beq
\gamma_{ij} \equiv\int d\br \delta n_i(\br) \delta n_j(\br) /
\tau'(\br) \, .
\label{eq:temperatures84}
\eeq 
\begin{figure}
  \centerline{\epsfig{file=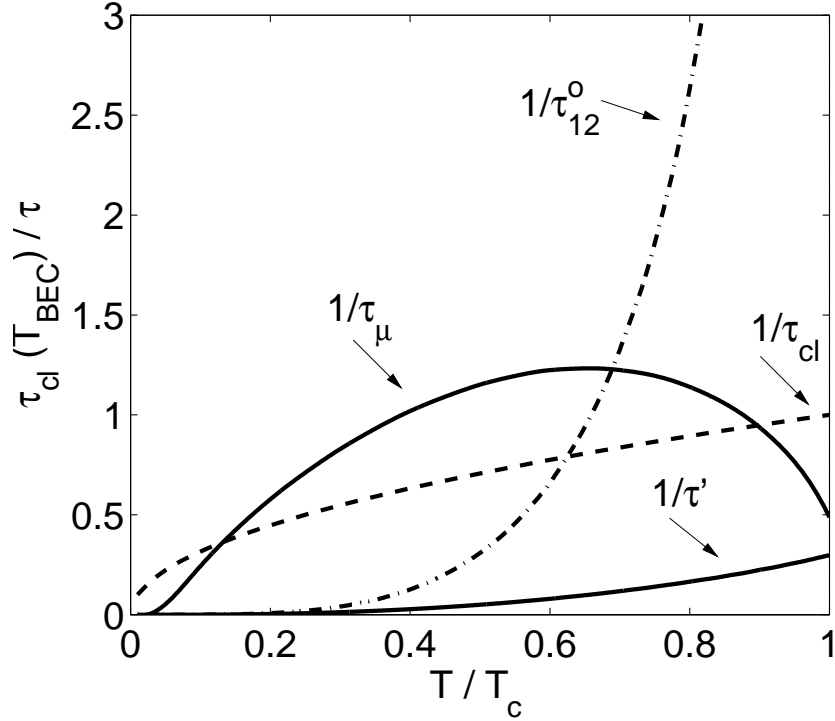,width=4.4in}}
\caption{Collision rates in a homogeneous Bose-condensed gas,
normalized to the collision time~(\ref{eqcoupled95}) in a classical 
gas at the BEC
transition temperature $\tau_{\rm{cl}}^{-1}(T_{\rm{BEC}})$. We have
taken $gn = 0.1 k_{\rm{B}} T_{\rm{BEC}}$, where $n$ is the total
density. See also Ref.~\protect\cite{Wilgrif}.} \label{fig:canberra3}
\end{figure}

Assuming the damping is small (our present discussion is for the 
collisionless
region), (\ref{eq:temperatures83}) is easily solved using 
perturbation theory by setting
$\gamma_{ij}=0$ for $j\neq i$. This gives the damped Stringari
frequency (to lowest order) $\Omega_i = \omega_i - i \Gamma_i$, with
\beq \Gamma_i \equiv {\gamma_{ii} \over 2} = {1 \over 2} \int d \br {
\delta n_i(\br)^2 \over \tau'(\br)}.
\label{eq:temperatures85}
\eeq 
This result for $\Gamma_i$ is reasonable, namely it involves a spatial 
average over $1/\tau'(\br)$ weighted with respect to the undamped
density fluctuations of the Stringari wave equation 
(\ref{eq:temperatures81}). Calculation shows that the effect of 
coupling to other modes ($\gamma_{ij}\neq 0$) is extremely
small.

Before discussing the trapped gas, it is useful to first apply our
theory to a homogeneous gas, where $\tau^\prime$ is independent of
position.  In this case, (\ref{eq:temperatures85}) simply reduces to 
$\Gamma_i = 1/2\tau'$. Although
our model  applies only to the collisionless
region, it is useful to compare the inter-component collision time 
for a uniform gas in
the collisionless and hydrodynamic regimes. In Section 
~\ref{sec:fluid}, we show  that the inter-component collision time 
$\tau_{\mu}$ in the
hydrodynamic region~\cite{Zarnikgrif} is given by 
$\tau_{\mu}=\sigma_H\tau'$, where the
temperature-dependent factor $\sigma_H$ depends on various 
thermodynamic
functions [$\sigma_H$ is given explicitly by (\ref{eqfluid165}) in 
Section~\ref{sec:coupled}].
In Fig.~\ref{fig:canberra3}, we compare $1/\tau^\prime$ and 
$1/\tau_{\mu}$ as
functions of $T$. We see that $\sigma_H$ dramatically alters the
inter-component relaxation rate $1/\tau_{\mu}$ appropriate to the
hydrodynamic regime, as compared to $1/\tau^\prime$ involved in the
collisionless regime. For completeness, in Fig.~\ref{fig:canberra3} 
we also plot the
often-used classical collision time  as well as
$\tau_{12}^0$ defined in (\ref{eq:temperatures77}).

We briefly discuss some numerical  calculations~\cite{Wilgrif} of the 
inter-component damping using the above formalism.   We  consider 
$^{87}$Rb atoms in a spherically symmetric trap with frequency $\nu_0 
= 10$ Hz and 
$N=2\times 10^6$. In the collisionless limit, we require $\omega_i
\tau_{\rm{cl}} \gg 1$. We obtain an upper limit on $1/\tau_{\rm{cl}}$ 
by
taking the density in the center of the trap $n(r=0)$, which gives
$1/\tau_{\rm{cl}} =8a^2N\omega_0^3 m /(\pi k_{\rm{B}}T)$. For the
parameters we use, $\omega_{10} \tau_{\rm{cl}} \approx 19$ (compared
to $\omega_{02} \tau_{\rm{cl}} \approx 20$ for the JILA
data~\cite{Jin97} and $\omega_{02} \tau_{\rm{cl}} \approx 2$ for the
MIT data~\cite{Stamper-Kurn98} on collective oscillations at finite 
$T$).

\begin{figure}
  \centerline{\epsfig{file=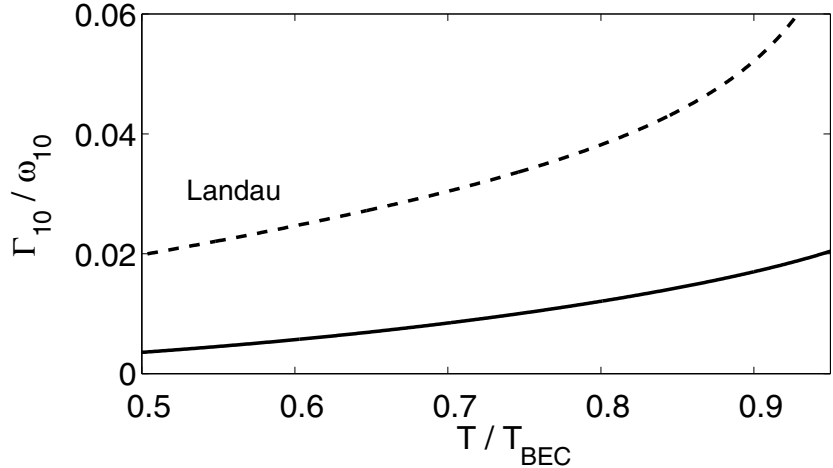,width=4.4in}}
\vspace{-1.8in}
\caption{Breathing mode damping rates vs. temperature. In these plots,
the damping rates are normalized to the mode
frequencies and we only plot up to $T=0.95T_{\rm{BEC}}$, above which
the Thomas-Fermi approximation will start to break down. We show
the inter-component damping rate given by (\ref{eq:temperatures80}) 
and (\ref{eq:temperatures85}) for the breathing mode $(n=1, l=0)$
of a spherically symmetric trap, as well as the Landau 
damping.~\protect\cite{Pitaevskii97}
For further details, see Ref.~\protect\cite{Wilgrif}.} 
\label{fig:canberra4}
\end{figure}

In Fig.~\ref{fig:canberra4} we plot the damping rate $\Gamma_{10}$ 
for the breathing
mode ($n=1$, $l=0$) as a function of temperature up
to $T=0.95 T_{\rm{BEC}}$, where $N_c\approx7\times 10^4$. At higher
temperatures, the Thomas-Fermi approximation will start to break down
and the mode frequencies become temperature
dependent.

The damping of condensate modes we have discussed in this section is 
due to the fact that the condensate is out of diffusive equilibrium 
with the thermal cloud. At finite $T$, the collective oscillations of 
the coupled condensate and non-condensate can be generally split into 
two classes.  One mode mainly involves  (out-of-phase) motion of the 
condensate and is the finite $T$ generalization of the analogous 
$T=0$ condensate mode.~\cite{Sstringari} For the same symmetry, there 
is another mode  which mainly involves the (in-phase) motion of the 
non-condensate thermal cloud.  This mode is naturally viewed as a 
generalization of the $T>T_{BEC}$ thermal cloud oscillation to 
temperatures below $T_{BEC}.$  These two types of modes (for a given 
symmetry) have been obtained~\cite{Zarnikgrif} in the 
collision-dominated hydrodynamic region, when the dynamics of the 
thermal cloud is fully allowed for.  Treating (to first 
approximation) the thermal cloud (the non-condensate component) as 
always in {\em static} equilibrium in this section means that our 
theory is only applicable to condensate oscillations which involve 
motions which are  out-of-phase with the thermal cloud.

As we discuss in Section~\ref{sec:coupled}, the possibility that the 
condensate may be out of diffusive equilibrium with the 
non-condensate is a general feature of the dynamics of trapped 
Bose-condensed gases.  In the hydrodynamic two-fluid region, it leads 
to a new relaxational mode.~\cite{Nikuni99}  This effect is missed in 
the well-known form of the two-fluid hydrodynamics developed by 
Landau~\cite{Land,Khalatnikov}, where one (implicitly) assumes that 
the normal and superfluid components are always in local diffusive 
equilibrium.  We will return to this question in 
Sections~\ref{sec:coupled} and \ref{sec:fluid}.

Of course, in addition to the inter-component relaxation discussed in 
this section, one {\em also} has Landau and Beliaev damping of 
condensate oscillations.~\cite{Huashigrif,Sgior,Fedichev98a}  These 
mechanisms are treated in more detail by Burnett in this book. Landau 
damping is briefly discussed in 
Section~\ref{sec:coupled}.

Clearly one can extend the formalism of this section to any problem 
based on the $T=0$ GP equation.  In particular, it can be used to 
discuss this kind of inter-component damping at finite $T$ of 
oscillations in two-component Bose gases,  vortex 
dynamics~\cite{Alfet,ALfett}, and Josephson oscillations between two 
traps.

\section{Dynamics of the coupled condensate and non-condensate at 
finite temperatures}
\label{sec:coupled}

In contrast to Section~\ref{sec:temperatures}, we now grapple with 
the dynamics of the thermal cloud.  As we noted in 
Section~\ref{sec:temperatures}, in dealing with this very complicated 
problem, we will be quite modest and treat the non-condensate using 
the simplest microscopic model which captures the important physics.  
We limit ourselves to finite $T$, where the non-condensate atoms can 
be described by the particle-like HF spectrum given by 
(\ref{eq:temperatures59}).
For trapped Bose gases, this spectrum is probably adequate down to 
quite low temperatures, for reasons discussed in an important paper 
by the Trento group.~\cite{Giorgini97} To extend our present analysis 
to very low temperatures is much more involved since then the 
excitations of the thermal cloud take on a collective aspect (ie, a 
Bogoliubov-type quasiparticle spectrum must be used).~\cite{Tomgrif}

The single-particle spectrum (\ref{eq:temperatures59}) is appropriate 
in the semi-classical limit, where we can use the single-particle 
distribution function $f({\bf p}, {\bf r}, t)$ given by the solution 
of a kinetic equation.
This procedure generalizes the approach of Boltzmann (1880's) for 
describing binary collisions in a classical gas.  Such a quantum 
Boltzmann equation for a trapped Bose-condensed gas at finite 
temperatures has been derived and extensively discussed by Zaremba, 
Nikuni and the author.~\cite{Zarnikgrif}   The conditions of validity 
are
\begin{equation}
k_BT\gg gn_{c}\ , \ k_BT\gg \hbar\omega_0 , 
\label{eqcoupled86}\end{equation}
where $\hbar\omega_0$ is the spacing of the energy levels of the 
harmonic trap. More general but less explicit discussions of kinetic 
equations are given in Refs.~\cite{Stoof99,Walser99}.  Related work 
of Gardiner and coworkers~\cite{Gardiner2000} is discussed in the 
lectures by Ballagh in this book.
The quantum kinetic equation we use is given by
\begin{eqnarray}
{\partial f({\bf p}, {\bf r}, t)\over\partial t} &+&{{\bf p}\over 
m}\cdot\mbox{\boldmath$\nabla$}_r  
f({\bf p}, {\bf r}, t) - \mbox{\boldmath$\nabla$}_r 
U({\bf r}, t) \cdot\mbox{\boldmath$\nabla$}_p f({\bf p}, {\bf r}, t) 
\nonumber \\
&=& C_{22}[f] + C_{12}[f].\label{eqcoupled87}\end{eqnarray}
The right hand side describes how binary collisions effect the value 
of the single-particle distribution function $ f({\bf p}, {\bf r}, 
t)$. The effective time-dependent HF potential $U({\bf r}, t)$ is 
defined in (\ref{eq:temperatures59}).   
The effect of collisions between {\em excited} atoms in the 
non-condensate is described by:
\begin{eqnarray}
C_{22}[f] &=& {2g^2\over (2\pi)^5}\int d{\bf p}_2\int d{\bf p}_3\int 
d{\bf p}_4\delta({\bf p}+{\bf p}_2 - {\bf p}_3 - {\bf p}_4)\nonumber 
\\
&\times&\delta\left({\tilde\varepsilon}_p+
{\tilde\varepsilon}_{p_2}-{\tilde\varepsilon}_{p_3}
-{\tilde\varepsilon}_{p_4}\right)\nonumber 
\\&\times&\left[(1+f)(1+f_2)f_3f_4 - 
ff_2(1+f_3)(1+f_4)\right].
\label{eqcoupled88}
\end{eqnarray}
We recall that creating a Boson gives a factor $(1 + f)$ and 
destroying a Boson gives $f.$  In the classical high temperature 
limit, $f\ll 1$ and the collision integral $C_{22}$ in 
(\ref{eqcoupled88}) becomes much simpler. 

In addition to $C_{22}$ collisions, we also have collisions which 
involve {\it one} condensate atom:
\begin{eqnarray}
C_{12}[f] &=& {2g^2\over (2\pi)^2}\int d{\bf p}_1 \int d{\bf p}_2 
\int d{\bf p}_3\delta (m{\bf v}_c +{\bf p}_1-{\bf p}_2-{\bf p}_3) 
\nonumber\\
&\times& \delta\left(\varepsilon_c 
+{\tilde\varepsilon}_{p1}-{\tilde\varepsilon}_{p2}
-{\tilde\varepsilon}_{p3}\right)\left[\delta\left({\bf 
p}-{\bf p}_1)-\delta({\bf p}-{\bf p}_2)-\delta({\bf p}-{\bf 
p}_3\right)\right]\nonumber \\
&\times&\left[n_c(1+f_1)f_2f_3 
-n_cf_1(1+f_2)(1+f_3)\right].
\label{eqcoupled89}
\end{eqnarray}
For convenience, we recall that the condensate atom has
\begin{eqnarray}
&{}&\mbox{energy:} \  \varepsilon_c = \mu_c + {1\over 2} mv_c^2 \ ; \ 
\mu_c = V_{ex} + gn_c +2g{\tilde n}\nonumber\\
&{}&\mbox{momentum:} \  {\bf p}_c = m{\bf v}_c 
\label{eqcoupled90}
\end{eqnarray}
We note the {\it key} difference between $C_{12}$ and $C_{22}$ 
collisions:
\begin{itemize}
\item $C_{22}$ and $C_{12}$ both conserve energy and momentum in 
collisions.
\item $C_{12}$ does {\it not} (but $C_{22}$ does) conserve the number 
of condensate atoms. $C_{12}$ describes how atoms are ``kicked'' in 
and out of the condensate.
\end{itemize}

It turns out in the generalized GP equation given in 
(\ref{eq:temperatures63}), the damping term $-iR({\bf r}, t)$ is 
closely related to $C_{12}[f]$.  This makes sense, since the $C_{12}$ 
collisions  modify the condensate described by $\Phi({\bf r}, t).$  
One easily may verify that [comparing (\ref{eq:temperatures66}) and 
(\ref{eqcoupled89})]
\begin{equation}
\Gamma_{12}[f, \Phi] =\int {d{\bf p}\over(2\pi)^3} C_{12}[f({\bf p}, 
{\bf r}, t)].\label{eqcoupled91}\end{equation}
\vskip 2pt
\noindent More precisely, the  three-field correlation function in 
the exact equation of motions [see (\ref{eq:temperatures56}) and 
(\ref{eq:temperatures58})] is related to $C_{12}[f]$ by
\begin{equation}
\int{d{\bf p}\over (2\pi)^3} C_{12}[f] = 2g\sqrt{n_c}\ 
Im\left<{\tilde\psi}^\dagger{\tilde\psi}{\tilde\psi}\right>.
\label{eqcoupled92}
\end{equation}

We have to solve the equation of motions for  $f({\bf p}, {\bf r}, 
t)$ and $\Phi({\bf r}, t)$ treating $C_{12}[f]$ very carefully. There 
will be an exchange of atoms between the ${\tilde n}({\bf r}, t)$ and 
$n_c({\bf r}, t)$ components through the $C_{12}$ collisions.  We can 
use these coupled equations for a variety of problems.  Recently 
these coupled equations have been used in two problems:
\begin{enumerate}
\item[(a)] Derivation of a generalized set of two-fluid hydrodynamic 
equations.~\cite{Zarnikgrif}
\item[(b)] Discussion of the rate of growth of a 
condensate~\cite{Bijlsma2000} due to a sudden quench in which the 
high energy spectrum of the thermal cloud distribution is suddenly 
removed.  This work has many points of contact with the formalism 
reviewed by Ballagh in this book for condensate 
growth.~\cite{Gardiner2000} 
\end{enumerate}
In this section, we will linearize these equations and consider the 
collective oscillations of the combined system composed of condensate 
and non-condensate.~\cite{Gardiner2000}  
It is useful to introduce {\it two} regimes to describe collective 
modes in interacting systems~\cite{Gorkov,chap1}:
\begin{enumerate}
\item[I.]Collisionless (produced by {\it mean fields})
\beq
\omega\tau_R\gg 1 \ \ \mbox{or}\ \ T\ll \tau_R \ 
\left(\omega\equiv{2\pi\over T}\right)
\label{eqcoupled93}\eeq
\item[II.] Hydrodynamic (produced by {\it collisions})

\beq\omega\tau_R\ll 1 \ \ \mbox{or}\ \ T\gg \tau_R,
\label{eqcoupled94}\eeq
\end{enumerate}
where $\tau_R$ is some appropriate relaxation time.
What should we use for $\tau_R$?  For a uniform classical gas, this 
is the collision time~\cite{Statistical,Nikuni99} 
\begin{equation}
{1\over\tau_c}=\sqrt 2{\tilde n}\sigma {\bar 
v}\label{eqcoupled95}\end{equation}
where
\begin{eqnarray*}
&{}&\sigma= 8\pi a^2 \ \mbox{(for Bose atoms)}; \ a = s\mbox{-wave 
scattering length}.\\
&{}&{\bar v} \simeq \ \mbox{average velocity of atoms}\ 
\sim\sqrt{k_BT\over m}.\\
&{}&{\tilde n}  =\  \mbox{density of excited atoms}.
\end{eqnarray*}
Even for a Bose-condensed gas, taking $\tau_R \sim\tau_c$ is a 
reasonable first estimate.
To get into the interesting hydrodynamic region $(\omega\tau_R\ll 
1)$, we need a small value of $\tau_R$, ie, a large  density ${\tilde 
n}$ or a large collision cross-section  $\sigma$ (perhaps by working 
near a Feshbach resonance).  We note that current BEC experiments 
deal with $N\simeq 10^6$ atoms or larger and thus the kind of density 
of thermal atoms needed to be in the hydrodynamic region now seem 
achievable.

Let us look at the kinetic equation (\ref{eqcoupled87}), writing it 
in the schematic form:
\begin{equation}{\hat{\cal L}} f = C_{22}[f] 
+C_{12}[f].\label{eqcoupled96}\end{equation}
In the collisionless region, we need only solve ${\hat{\cal L}} f = 
0.$
For comparison with the collision-dominated hydrodynamic domain, we 
briefly discuss some features of this collisionless domain, limiting 
ourselves to a {\it uniform} gas for simplicity.  This enables us to 
understand how Landau and Beliaev damping arise from the 
collisionless Boltzmann equation, in contrast to the inter-component 
collisional damping (discussed in Section \ref{sec:temperatures}) 
which has its origin in the $C_{12}[f, \Phi]$ collision integral.

We consider linear response theory~\cite{chap1,Kadbaym} for the 
deviations of $f(\bp, \br, t)$ from thermal equilibrium
\beq
f(\bp, \br, t) = f_0(\bp, \br) +\delta f(\bp, \br, 
t).\label{eqcoupled97}\eeq
For a uniform normal Bose gas to first order in $\delta f,$ the 
collisionless kinetic equation reduces to
\beq {\partial\delta f\over\partial t}+{\bp\over 
m}\cdot\mbox{\boldmath$\nabla$}_r\delta 
f-\mbox{\boldmath$\nabla$}_r[2g\delta n(\br, t)]+\delta U(\br, 
t)]\cdot\mbox{\boldmath$\nabla$}_p f_0(\bp)=0.\label{eqcoupled98}\eeq
Assuming $\delta U(\br, t)=\delta U_{k\omega}e^{i({\bf 
k}\cdot\br-\omega t)}$, then we have $\delta f(\bp, \br, t) = \delta 
f_{k\omega}(\bp)e^{i({\bf k}\cdot\br-\omega t)}$, where $\delta 
f_{k\omega}(\bp)$ satisfies
\beq -i\omega\delta f(\bp) + i{\bp\cdot{\bf k}\over m}\delta 
f(\bp)-[2g\delta n_{k\omega}+\delta U_{k\omega}]i{\bf 
k}\cdot\mbox{\boldmath$\nabla$}_p f_0(\bp) = 0\ 
,\label{eqcoupled99}\eeq
where
\beq f_0({\bf p})={1\over e^{\beta({p^2\over 2m}+gn_{0}-\mu_{0})}-1} 
\equiv f_0(\epsilon_p)\label{eqcoupled100}\eeq
and \beq\delta n_{k\omega}\equiv\int{d\bp\over(2\pi)^3} \delta 
f_{k\omega}(\bp).\label{eqcoupled101}\eeq
Solving (\ref{eqcoupled99}), one finds
\beq\delta n_{k\omega}=-\int{d\bp\over (2\pi)^3} {\bp\cdot{\bf 
k}\over m}{\partial f_0(\epsilon_p)\over\partial\epsilon_p} 
{[2g\delta n_{k\omega}+\delta U_{k\omega}]\over \omega - 
{\bp\cdot{\bf k}\over m}}\label{eqcoupled102}\eeq
or
\beq\delta n_{k\omega}={\chi^0_{nn}({\bf k},\omega)\over 
1-2g\chi^0_{nn}(k,\omega)}\delta U_{k\omega} \equiv\chi_{nn}({\bf 
k},\omega)\delta U_{k\omega}.\label{eqcoupled103}\eeq
Here
\beq\chi^0_{nn}({\bf k}, \omega)=-\int{d\bp\over 
(2\pi)^3}{\bp\cdot{\bf k}\over m}{\partial 
f_0(\epsilon_p)\over\partial\epsilon_p} {1\over\omega-{\bp\cdot{\bf 
k}\over m}}\label{eqcoupled104}\eeq
is the $k\to 0$ limit of the density response function of a 
non-interacting uniform Bose gas.

This shows how one can use the collisionless kinetic equation to find 
the density response function $\chi_{nn}({\bf k}, \omega)$ of a 
uniform Bose gas.  The density fluctuations are given by the poles of 
$\chi_{nn}({\bf k}, \omega) $ in (\ref{eqcoupled103}).  In this RPA 
with exchange, the solutions of
\beq 1-2g\chi_{nn}^0({\bf k}, \omega) = 0\label{eqcoupled105}\eeq
are called ``zero sound'' modes (following Landau's terminology in 
the analogous calculation for an interacting Fermi 
gas~\cite{chap1}).  It is clear that the solutions $\omega(k)$ of 
(\ref{eqcoupled105}) will be Landau damped, with a width related to
\bea &-&2g Im \chi_{nn}^0 ({\bf k}, \omega + i0^+)\nonumber\\
&=&2g\pi\int{d\bp\over(2\pi)^3}[f_0(\epsilon_p)-f_0(\epsilon_{p+k})]
\delta[\omega(k)-(\epsilon_{p+k}-\epsilon_p)].
\label{eqcoupled106}
\eea
Clearly this damping will come from the zero sound mode absorbing a 
thermal excitation of energy $\epsilon_p$ to create a thermal 
excitation $\epsilon_{p+k}$, 
\beq\hbar\omega(k)=\epsilon_{p+k}-\epsilon_p\ 
..\label{eqcoupled107}
\eeq

We might also remark that if this linear response calculation is 
extended to deal with a Bose-condensed gas, one finds a new damping 
mechanism because now $\chi^0_{nn}({\bf k},\omega)$ describes 
non-interacting Bogoliubov excitations $E_p$.  In addition to the 
contributions of the kind given by (\ref{eqcoupled107}), one finds 
$\chi_{nn}^0({\bf k}, \omega)$ now has additional poles at
\beq \hbar\omega(k) = E_{p+k}+E_p.\label{eqcoupled108}\eeq
These give rise to ``Beliaev'' damping, where the zero sound mode can 
decay into two Bogoliubov excitations.  In the limit of $T\to 0$, 
Beliaev damping dominates over Landau damping.  It was first 
calculated at $T=0$ by Beliaev~\cite{Bel2rev} in 1957 in his classic 
study of a weakly interacting Bose gas.

Several recent papers have discussed Landau and Beliaev damping of 
collective modes in a trapped Bose 
gas.~\cite{Guilleumas2000,Fedichev98a,Sgior,Morgan2000,Reidl}  This 
problem has been  nicely formulated directly in terms of response 
functions by Minguzzi and Tosi~\cite{Mingtosi} in conjunction with 
mean fields produced by the condensate and non-condensate density 
fluctuations.  Giorgini~\cite{Sgior} has given a physically appealing 
discussion working directly with the coupled mean field equations of 
motion for $\Phi(\br, t), {\tilde n}(\br, t) $ and ${\tilde m}(\br, 
t).$  
The approach of the Oxford group is reviewed in Burnett's lectures in 
this book.

In this connection, a simple treatment of Landau damping of 
condensate collective modes can be given starting from our 
generalized GP equation in (\ref{eq:temperatures63}).  In 
Section~\ref{sec:temperatures}, we ignored the fluctuations in the HF 
term $2g{\tilde n}(\br, t)$ and concentrated on the damping 
associated with the $-iR(\br, t)$ term in  
(\ref{eq:temperatures69}).  A simple way of including the 
$\delta{\tilde n}(\br, t)$ fluctuations induced by the mean field 
associated with the condensate fluctuations is to use (see 
Refs.~\cite{Mingtosi,Wilgrif})
\beq\delta{\tilde n}(\br, \omega)=\int d\br^\prime \chi^0_{{\tilde 
n}{\tilde n}}(\br, \br^\prime, \omega) 2g\delta n_c(\br^\prime, 
\omega).\label{eqcoupled109}\eeq
Here $ \chi^0_{{\tilde n}{\tilde n}}$ is the density response 
function for a non-interacting gas of atoms with the equilibrium HF 
spectrum given by (\ref{eq:temperatures69}) and the chemical 
potential ${\tilde\mu}_0$ is equal to $\mu_{c0}$ in 
(\ref{eq:temperatures68}). For a uniform Bose gas, the condensate 
collective modes given by (\ref{eq:temperatures63}) are found to be  
\beq \omega^2=c_0^2k^2[1+4g\chi^0_{{\tilde n}{\tilde n}}({\bf k}, 
\omega)] -{i\omega\over \tau^\prime}\ .\label{eqcoupled110}\eeq Here 
$c_0=\sqrt{gn_{c0}/m}$ is the Bogoliubov phonon velocity [see 
(\ref{eq:dynamics28})] and $\tau^\prime$ is defined in 
(\ref{eq:temperatures80}).  In the long wavelength limit, one 
finds~\cite{Wilgrif,Reidl,Szepfalushy74}
\beq\lim_{k\to 0} \,-Im \,\chi^0_{{\tilde n}{\tilde n}} ({\bf k}, 
\omega = c_0k)={mk_BT\over 2\pi c_0}{1\over 
3},\label{eqcoupled111}\eeq
where the $1/3$ factor arises from the exchange term in the 
single-particle self-energy.
Thus one obtains \beq\omega=c_0k-i(\Gamma_L+{1\over 
2\tau^\prime}),\label{eqcoupled112}\eeq
where the Landau damping is given by (RPA with exchange)
\beq \Gamma_L={4\over 3}\,a\, k_BTk.\label{eqcoupled113}\eeq
The exact result for $\Gamma_L$ obtained from a careful evaluation of 
the Beliaev self-energies~\cite{Huashigrif,Fedichev98a,Pitaevskii97} 
at finite $T$ is the same as (\ref{eqcoupled113}), apart from a 
slightly smaller numerical coefficient ($4/3$ is replaced by 
$3\pi/8)$.  This difference is due to our neglect of the fluctuations 
associated with the anomalous pair density ${\tilde m}(\br, t)$.  
Such terms~\cite{Sgior} are in the exact equation for $\Phi(\br, t)$ 
given in (\ref{eq:temperatures53}) but were neglected in deriving 
(\ref{eq:temperatures63}).

To conclude this discussion of different kinds of damping in the 
collisionless limit, we can also include the effect of the collision 
terms in (\ref{eqcoupled87}) as a perturbative correction to the 
solution determined by ${\hat{\cal L}} f = 0.$  For a Bose gas {\it 
above} $T_{BEC}$, one might use the simple ``relaxation time'' 
approximation~\cite{Kavoulakis98}\beq C_{22}[f] \simeq 
-(f-f_0)/\tau_{22}=-\delta f/\tau_{22}\ .\label{eqcoupled114}\eeq
The inclusion of such a term in (\ref{eqcoupled98}) simply involves 
the change
\beq{\partial\delta f\over\partial t}\to{\partial\delta 
f\over\partial t}+{\delta f\over\tau_{22}},\label{eqcoupled115}\eeq
and hence one finds (\ref{eqcoupled103}) again, with the replacement 
$\omega\to\omega+i/\tau_{22}$ in the denominator of 
(\ref{eqcoupled104}).  The collision time $\tau_{22}$ is the analogue 
of the classical expression given in (\ref{eqcoupled95}).  One sees 
that including $C_{22}[f]$ in the simple relaxation time 
approximation (\ref{eqcoupled114}) only slightly alters the Landau 
damping given by (\ref{eqcoupled106}).  This is quite different from 
the contribution arising from $C_{12}[f]$ in so far as this enters 
directly into the generalized GP equation (see 
Section~\ref{sec:temperatures}).

We now leave the discussion of the collisionless domain and turn to 
the collision-dominated hydrodynamic region.  In this case, the 
collisions between the non-condensate atom are assumed to be 
sufficiently rapid that the form of $f(\bp, \br, t)$ is largely 
determined by the requirement that
\beq C_{22}[f, \Phi]=0.\label{eqcoupled116}\eeq
That is, the collisions are so strong they force the system to be in 
{\it local} equilibrium.  The unique solution (denoted by ${\tilde 
f})$ of the equation (\ref{eqcoupled116}) is well-known to be given by
\begin{equation}
{\tilde f}({\bf p}, {\bf r}, t) = {1\over e^{\beta[{({\bf p}-m{\bf 
v}_n)^2\over 2m}+U({\bf r}, t) - {\tilde\mu}(r, 
t)]}-1},\label{eqcoupled117}\end{equation}
where ${\bf v}_n$ is the average local velocity and ${\tilde\mu}$ is 
the local chemical potential of the thermal atoms.
This local equilibrium Bose distribution involves the local variables 
$\beta, {\bf v}_n, {\tilde\mu}$ and $U$, all of which depend on 
$({\bf r}, t)$.
Why must ${\tilde f}$ have the form in (\ref{eqcoupled117})?  To 
satisfy $C_{22}[f_1] = 0,$ we must have [see (\ref{eqcoupled88})]
\begin{equation}
(1+f_1)(1+f_2)f_3f_4 - 
f_1f_2(1+f_3)(1+f_4)=0,\label{eqcoupled118}\end{equation}
and this requires that
$f$ be given by the Bose distribution (\ref{eqcoupled117}).  Here we 
have used the fact that a Bose distribution satisfies \begin{equation}
f(x)\equiv{1\over e^x-1}  = 
-[f(-x)+1]\label{eqcoupled119}\end{equation}
and energy and momentum conservation
\begin{equation}
\left.\begin{array}{cc}
&{\bf p}_1 + {\bf p}_2 = {\bf p}_3 + {\bf p}_4 \nonumber \\
&{}\nonumber\\
& {\tilde\varepsilon}_{p_1}+{\tilde\varepsilon}_{p_2}=
{\tilde\varepsilon}_{p_3}+
{\tilde\varepsilon}_{p_4}\end{array}\right\}.
\label{eqcoupled120}
\end{equation}

However, while the fact that ${\tilde f}$ is given by the local 
equilibrium Bose distribution in $(\ref{eqcoupled117})$ ensures that 
$C_{22}[{\tilde f}]=0, $ one finds~\cite{Zarnikgrif,Nikuni99} that 
$C_{12}[{\tilde f}] \ne 0.$  More precisely, the term in 
(\ref{eqcoupled89}) reduces to \begin{eqnarray}
&{}&\left[(1+{\tilde f}_1){\tilde f}_2{\tilde f}_3 - {\tilde 
f}_1(1+{\tilde f_2})(1+{\tilde f_3})\right] \nonumber\\
&{}&\propto \left[e^{-\beta[{\tilde\mu}-\mu_c-{1\over 2}m({\bf 
v}_n-{\bf v}_c)^2]}-1\right](1+{\tilde f}_1){\tilde f}_2{\tilde 
f}_3,\label{eqcoupled121}\end{eqnarray}
using energy and momentum conservation.  
The expression in the square bracket in (\ref{eqcoupled121}) only 
vanishes if the condensate and non-condensate are in {\it diffusive} 
equilibrium, which requires that the chemical potentials must be equal
\begin{equation}
{\tilde\mu}=\mu_c+{1\over 2}m({\bf v}_n-{\bf 
v}_c)^2.\label{eqcoupled122}\end{equation}
When we {\it perturb} the system, this may not be true, ie, the two 
components may be out of diffusive equilibrium.

We can now derive hydrodynamic equations for the non-condensate by 
taking moments of the Boltzmann equation, the standard procedure used 
in classical gases.~\cite{Statistical} The first moment gives a 
continuity equation with a source term:
\begin{eqnarray}
\int d{\bf p}\left\{{\cal L}{\tilde f} = C_{12}[{\tilde 
f}]\right\}\rightarrow {\partial{\tilde n}\over\partial t} = 
-\mbox{\boldmath$\nabla$}\cdot({\tilde n}{\bf v}_n)+
\Gamma_{12}[{\tilde f}], \label{eqcoupled123}\end{eqnarray}
where we have used the fact that $C_{22}[{\tilde f}]=0$ and 
\begin{eqnarray}
{\tilde n} &\equiv& \int{d{\bf p}\over (2\pi)^3} {\tilde f}({\bf p}, 
{\bf r}, t)\nonumber \\
{\tilde n}{\bf v}_n &\equiv& \int{d{\bf p}\over (2\pi)^3} {{\bf 
p}\over m}{\tilde f}({\bf p}, {\bf r}, t)\nonumber\\
\Gamma_{12}[{\tilde f}, \Phi]&\equiv&\int{d{\bf p}\over(2\pi)^3} 
C_{12}[{\tilde f}, \Phi].\label{eqcoupled124}\end{eqnarray}
More explicitly, we find [compare with (\ref{eq:temperatures71})]
\begin{eqnarray}
\Gamma_{12}[{\tilde f}]&=& {2g^2n_c\over(2\pi)^5} 
[e^{-\beta[{\tilde\mu}-\mu_c-{1\over 2}m({\bf v}_n-{\bf 
v}_c)^2]}-1]\nonumber\\
&\times&\int d{\bf p}_1\int d{\bf p}_2\int d{\bf p}_3 \delta(m{\bf 
v}_c +{\bf p}_1 - {\bf p}_2 - {\bf p}_3) \nonumber \\
&\times&\delta(\varepsilon_c+{\tilde\varepsilon}_1-
{\tilde\varepsilon}_2-{\tilde\varepsilon}_3)(1+{\tilde 
f}_1){\tilde f}_2{\tilde f}_3 \nonumber \\
&\equiv&{n_c\over\tau_{12}}\left[e^{-\beta[{\tilde\mu}-
\mu_c-{1\over 2}m({\bf v}_n-{\bf 
v}_c)^2]}-1\right].
\label{eqcoupled125}
\end{eqnarray}
We note that $\tau_{12}$ is a collision time which describes the 
$C_{12}$ collisions between the condensate and non-condensate atoms.  
Adding (\ref{eqcoupled123}) to the continuity equation in 
(\ref{eq:temperatures60}) 
\begin{equation}
{\partial n_c\over\partial t} = 
-\mbox{\boldmath$\nabla$}\cdot(n_c{\bf v}_c)-\Gamma_{12}[{\tilde f}, 
\Phi], \label{eqcoupled126}\end{equation}
we see that the source term $\Gamma_{12}$ cancels out to give
\begin{equation}{\partial(n_c+{\tilde n})\over\partial 
t}=-\mbox{\boldmath$\nabla$}\cdot(n_c{\bf v}_c+{\tilde n}{\bf 
v}_n).
\label{eqcoupled127}
\end{equation}
Thus our theory gives the correct continuity equation for the total 
local density $n=n_c+{\tilde n}.$

Similarly, one finds
\begin{eqnarray}
&&\int d{\bf p}{\bf p}\left\{{\hat{\cal L}}{\tilde f}\right.= 
\left.C_{12}[{\tilde f}]\right\}\rightarrow m{\tilde 
n}\left({\partial{\bf v}_n\over\partial t}+{1\over 
2}\mbox{\boldmath$\nabla$}{\bf v}^2_n\right)\nonumber\\
&& \ \ \ \ \  = -\mbox{\boldmath$\nabla$}{\tilde P}({\bf r}, 
t)-{\tilde n}\nabla U({\bf r}, t) -m({\bf v}_n-{\bf 
v}_c)\Gamma_{12}[{\tilde f}], \label{eqcoupled128}\end{eqnarray}
where the kinetic pressure is given  by
\begin{equation}
{\tilde P}({\bf r}, t) = {m\over 3}\int {d{\bf p}\over (2\pi)^3} 
({\bf p}-m{\bf v}_n)^2{\tilde f}({\bf p}, {\bf r}, 
t).\label{eqcoupled129}\end{equation}
Finally, the second moment gives
\begin{eqnarray}
&&\int  d{\bf p} p^2\left\{{\cal L}{\tilde f} = C_{12} [{\tilde 
f}]\right\}\rightarrow{\partial{\tilde P}\over\partial 
t}+\mbox{\boldmath$\nabla$}\cdot({\tilde P}{\bf v}_n)\nonumber \\
&&\  = -{2\over 3}{\tilde P}\mbox{\boldmath$\nabla$}\cdot{\bf 
v}_n+{2\over 3}\left[\mu_c +{1\over 2}m({\bf v}_n-{\bf 
v}_c)^2-U\right]\Gamma_{12}[{\tilde 
f}].\label{eqcoupled130}\end{eqnarray}
The detailed derivation of these equations is not important here.  It 
involves straightforward manipulations using the explicit form of 
${\tilde f}$ in (\ref{eqcoupled117}).  

The hydrodynamic equations (\ref{eqcoupled123}), (\ref{eqcoupled128}) 
and (\ref{eqcoupled130}) describe the non-condensate in terms of 
three new ``coarse-grained'' local variables:
\begin{displaymath}{\tilde n}({\bf r}, t), {\bf v}_n({\bf r}, t) \ 
\mbox{and}\ {\tilde P}({\bf r}, t).\end{displaymath}
These are coupled to the two additional local variables which 
describe the condensate:
\begin{displaymath}
n_c({\bf r}, t), \ {\bf v}_c({\bf r}, t). \end{displaymath}
We note that the two condensate equations of motion given by 
(\ref{eq:temperatures60}) are always ``hydrodynamic'' in form.  In 
contrast, it is only in the collision-dominated region that the 
non-condensate dynamics can {\em also} be described in terms of a few 
collective variables.  
Both components exhibit {\it coupled}, {\it coherent} collective 
motions at the same frequency.  This is the essence of two-fluid 
superfluid behaviour, familiar in liquid $^4$He studies but still an 
unexplored frontier in trapped Bose gases. 

We will now discuss the linearized version of our condensate and 
non-condensate equations for local equilibrium as given by 
(\ref{eq:temperatures60}), (\ref{eqcoupled122}), (\ref{eqcoupled128}) 
and (\ref{eqcoupled130}).  We work to first order in the fluctuations 
around static equilibrium, 
\bea{\tilde n}&=&{\tilde n}_0+\delta{\tilde n}, \bv_n = \delta\bv_n, 
\ {\tilde P}={\tilde P}_0+\delta{\tilde P}\nonumber\\
n_c&=&n_{c0}+\delta n_c, \bv_c=\delta\bv_c\ .\label{eqcoupled131}\eea
What is {\it new} about the two-fluid hydrodynamic equations derived 
above is the role of the source term $\Gamma_{12}[{\tilde f}].$  In a 
linearized theory expanded around the {\it static} equilibrium Bose 
distribution $f_0$
 (where $\Gamma_{12}[f_0, \Phi_0]$ vanishes),  one finds 
\begin{equation}
\Gamma_{12}[{\tilde f}, \Phi]=\delta\Gamma_{12}[{\tilde 
f},\Phi]=-{\beta_0 
n_{c0}\over\tau^0_{12}}\delta\mu_{diff},
\label{eqcoupled132}
\end{equation}

where
 \beq \mu_{diff}(\br, t)\equiv{\tilde \mu}
 (\br, t)-\mu_c(\br, t). \label{eqcoupled133}
\eeq 
Here $\tau^0_{12}(\br)$ is the $C_{12}$ 
collision time defined in (\ref{eqcoupled125}) with both components 
being in static equilibrium (see also (\ref{eq:temperatures77})).  
The linearized coupled hydrodynamic equations for the two components 
are given by the ZGN$^\prime$ equations~\cite{Zarnikgrif,Nikuni99}:
\bea{\partial\delta n_c\over\partial t}&=& 
-\mbox{\boldmath$\nabla$}\cdot(n_{c0}\delta\bv_c) 
-\delta\Gamma_{12}\nonumber\\
m{\partial\delta \bv_c\over\partial t}&=& 
-\mbox{\boldmath$\nabla$}\delta\mu_c\label{eqcoupled134}\eea
and
\bea {\partial\delta {\tilde n}\over\partial t}&=& 
-\mbox{\boldmath$\nabla$}\cdot({\tilde n}_0\delta\bv_n) 
+\delta\Gamma_{12}\nonumber\\
m{\tilde n}_0{\partial\delta \bv_n\over\partial t}&=& 
-\mbox{\boldmath$\nabla$}\delta{\tilde P}-\delta{\tilde 
n}\mbox{\boldmath$\nabla$} U_0(\br)-2g{\tilde 
n}_0[\mbox{\boldmath$\nabla$}\delta 
n_c+\mbox{\boldmath$\nabla$}\delta{\tilde n}]\nonumber\\
{\partial\delta {\tilde P}\over\partial t}&=& -{5\over 
3}\mbox{\boldmath$\nabla$}\cdot({\tilde P}_0\delta\bv_n)+{2\over 
3}\delta{\bv_n}\cdot\mbox{\boldmath$\nabla$}{\tilde P}_0-{2\over 
3}gn_{c0}\,\delta\Gamma_{12}\, ,\label{eqcoupled135}\eea
with $\delta\Gamma_{12}$ given by (\ref{eqcoupled132}) and (within 
the TF approximation)
\beq\delta\mu_c=g\delta n_c+2g\delta{\tilde 
n}.\label{eqcoupled136}\eeq
We note that to lowest order, $\delta\Gamma_{12}$ does not appear in 
the second equation in (\ref{eqcoupled135}).  In addition, in the 
last term in the third equation in (\ref{eqcoupled135}), we have used 
$\mu_{c0}-U_0=-gn_{c0}$.  The condensate couples into the 
non-condensate equations in 
(\ref{eqcoupled135}) directly via the mean field terms $g\delta n_c$ 
and, indirectly, through $\delta\Gamma_{12}$ (see 
Section~\ref{sec:fluid}).

Despite appearances, this coupled set of hydrodynamic equations form 
a closed set.  We have 9 scalar equations in (\ref{eqcoupled134}) and 
(\ref{eqcoupled135}), while they appear to involve 10 fluctuating 
variables
\beq \delta n_c, \delta\bv_c;\ \delta{\tilde n}, \delta\bv_n;\ 
\delta{\tilde P}, \delta\mu_{diff}.\label{eqcoupled137}\eeq
However one can show~\cite{Zarnikgrif} that $\delta\mu_{diff}$ can be 
written as a linear combination of the variables $\delta{\tilde P}, 
\delta{\tilde n}$ and $\delta n_c$ and thus we are in fact left with 
only 9 variables. In Section~\ref{sec:fluid}, we prove that 
(\ref{eqcoupled134}) and (\ref{eqcoupled135}) are equivalent to the 
two-fluid hydrodynamic equations of Landau when applied to a trapped 
Bose gas, but only when the condensate and non-condensate are in 
diffusive local equilibrium, ie, $\mu_c(\br, t) ={\tilde\mu}(\br, 
t).$  It will turn out that the rate at which $\delta\mu_{diff}(\br, 
t)$ relaxes to zero is controlled by a relaxation time $\tau_\mu$ 
related to the $C_{12}$ collisions.  The Landau two-fluid 
hydrodynamics~\cite{Khalatnikov} is only valid for low frequency 
oscillations which satisfy $\omega\tau_\mu\ll 1$, as we discuss in 
Section~\ref{sec:fluid}.

\section{Two-fluid hydrodynamics in Bose liquids and gases}
\label{sec:fluid}

To put our new two-fluid equations derived in 
Section~\ref{sec:coupled} into some sort of  context, we first 
briefly review the more general theory developed by Landau in 1941 to 
explain superfluidity in liquid $^4$He.

The original discovery of superfluidity in liquid $^4$He is 
associated with the famous 1938 papers of Kapitza in Moscow and Allen 
and Misener at Cambridge.  (I cannot resist remarking that Allen and 
Misener had been graduate students at the University of Toronto where 
they carried out (with Burton) the pioneering studies on vanishing 
viscosity in the period 1935-1937).  These and subsequent experiments 
in the next few years~\cite{classic} showed that superfluid $^4$He 
could exhibit very bizarre behaviour compared to ordinary liquids. 
This led to the development of a two-fluid theory of the hydrodynamic 
behaviour of liquid $^4$He by Landau (1941).  An earlier but less 
complete version of Landau's hydrodynamic equations was developed by 
Tisza in the period 1938-40.  For further discussion of this early 
history, I refer to a recent article of mine.~\cite{Flondon} 

In this early work, superfluidity (the term was coined in 1938 by 
Kapitza) was entirely associated with the {\em relative} motion of 
the normal fluid and the superfluid components under a variety of 
conditions.~\cite{classic} The main point was that while the normal 
fluid exhibited finite viscosity and thermal conductivity typical of 
ordinary fluid, the superfluid component (which  exhibited 
irrotational flow) did not.  In more recent times, the aspect of 
superfluidity which has been emphasized (see Ch.4 of 
Ref.~\cite{Gorkov} and Leggett's lectures in this book, for example) 
are those most directly tied to the fact that the superfluid velocity 
is associated with the gradient of the phase of the macroscopic 
wavefunction $\Phi(\br, t).$  Given that this point of view is indeed 
more fundamental, it is still crucial to understand why superfluidity 
persists {\it even} in the presence of a dissipative normal fluid.  
This question can be addressed using two-fluid hydrodynamic equations.

In essence, Landau developed his generic two-fluid hydrodynamics by 
generalizing the standard theory of classical hydrodynamics to 
include the equations of motion for a new ``superfluid'' degree of 
freedom.  We recall that classical fluid dynamics was developed well 
before one knew about the existence of atoms.  Since the work of 
Maxwell and Boltzmann in the 1880's, we know the ``coarse-grained'' 
hydrodynamic description of a fluid in terms of a few quantities like 
$n(\br, t)$ and $\bv(\br, t)$ is only valid when the collisions 
between atoms are strong enough to produce local 
equilibrium.~\cite{Statistical} It only describes low frequency 
phenomena, where the condition in (\ref{eqcoupled94}) is satisfied.

In his 1941 paper, Landau did not connect the superfluid component 
with the motion of a ``Bose condensate.''  Indeed, he rejected the 
efforts by Tisza and F. London~\cite{Flondon} to use a Bose-condensed 
gas to get some insight into superfluid $^4$He.  However, since the 
period 1957-1965, Landau's superfluid degree of freedom has been 
understood microscopically in terms of the complex order parameter 
$\Phi(\br, t).$  As noted earlier, the superfluid velocity field 
$\bv_s(\br, t)$ is related to the gradient of the phase of $\Phi(\br, 
t)$, as given by (\ref{eq:dynamics19}).  In the same period, it was 
also realized that there are two distinct kinds of quantum fluids:
\begin{enumerate}
\item[(a)] Bose fluids (associated with a Bose condensate 
wavefunction $\Phi$)
\item[(b)] Normal Fermi fluids (associated with the key role of a 
Fermi surface).
\end{enumerate}
These two kinds of quantum fluids were magnificently described in two 
books~\cite{chap1,Gorkov} by Nozi\`eres and Pines written around 
1965, although the one on superfluid Bose liquids was only published 
in 1990 (it was in wide circulation as a preprint before then).  I 
think the clearest account of the connection between $\Phi(\br, t)$ 
and superfluidity in Bose fluids is still the discussion given in 
Chapters 4 and 5 of Vol II by Nozi\`eres and Pines.~\cite{Gorkov}  I 
highly recommend it to everyone in the BEC field.  

Landau's pioneering work on superfluid $^4$He has two separate 
aspects which are logically distinct but sometimes confused with each 
other:\begin{enumerate}
\item[(a)] The two-fluid equations describing hydrodynamic 
behaviour.  These equations are generic and apply (under certain 
conditions) to trapped Bose gases as well as to superfluid $^4$He.  
In the period 1947-1950, Landau and Khalatnikov extended these 
equations to include hydrodynamic damping of the normal fluid 
(described by various kinds of viscosities, thermal conductivity, 
etc).  This work is all described in the classic 1965 monograph by 
Khalatnikov.~\cite{Khalatnikov}
\item[(b)] A ``microscopic'' theory of the elementary excitations 
describing the normal fluid of liquid $^4$He - the famous phonon - 
roton spectrum.    Within his picture of a weakly interacting gas of 
{\em quasiparticles}, Landau and coworkers could calculate the 
thermodynamic and transport properties of superfluid $^4$He.  These 
quantities enter into the expressions given by the two-fluid 
equations for the first and second sound modes (velocity and 
damping). Of course, the roton part of the spectrum is not valid for 
a dilute Bose gas.\end{enumerate}

As we have noted, Landau's original formulation of his two-fluid 
hydrodynamic equations was phenomenological in that the superfluid 
component was not given an explicit microscopic basis.  The first 
derivation of the Landau hydrodynamic equations starting simply from 
the existence of the macroscopic order parameter $\Phi(\br, t)$ was 
given by Bogoliubov in 1963.~\cite{Bogg}  This derivation built on 
Bogoliubov's earlier work on deriving hydrodynamic equations for 
classical liquids without going through the intermediate stage of 
using Boltzmann-like kinetic equations.  While Bogoliubov's 
derivation is often viewed as being quite general, buried in his 
complex analysis is the assumption that the normal fluid and the 
superfluid are in local equilibrium with each other.  Even today, 
probably the definitive account which formulates the various levels 
of theory for Bose superfluids is the classic paper by Hohenberg and 
Martin~\cite{Hohmar} published in 1965.  It clearly shows (to the 
patient reader!) the central unifying role of $\Phi(\br, t)$, 
summarizes the collisionless and hydrodynamic domains, and finally 
gives criteria for developing and judging various approximation 
schemes using Green's function techniques.

With the preceding discussion as a preamble, I will now summarize the 
linearized two-fluid hydrodynamic equations of Landau, following the 
approach given in Ch. 7 of Ref.~\cite{Gorkov}.  As noted, these 
equations are valid for both superfluid liquids as well as gases.  
The differences come in only at the last stage when we evaluate the 
thermodynamic coefficients appearing in these equations, using the 
appropriate quasiparticle excitation 
spectrum.~\cite{Griffin97b,Leeyang,Shenoy}  The first two 
(linearized) Landau equations are familiar from ordinary fluid 
dynamics
\begin{equation}
{\partial\delta n\over\partial t} + 
\mbox{\boldmath$\nabla$}\cdot\delta{\bf j} = 
0\label{eqfluid138}\end{equation}

\beq m{\partial\delta {\bf j}\over\partial t} = 
-\mbox{\boldmath$\nabla$}\delta P-\delta 
n\mbox{\boldmath$\nabla$}V_{ex},\label{eqfluid139}\eeq
where we have included the effect of an external potential.  Landau's 
work incorporated two components and hence the total mass density and 
mass current fluctuations are given by
\beq\delta \rho\equiv m\delta n = 
\delta\rho_s+\delta\rho_n\label{eqfluid140}\eeq
\beq m\delta{\bf j} \equiv\rho_{s0}\,\delta\bv_s 
+\rho_{n0}\,\delta\bv_n.\label{eqfluid141}\eeq

The new superfluid component was argued to only exhibit pure 
potential (irrotational) flow and carry no entropy. Thus Landau's 
final two hydrodynamic equations were new,
\beq
m{\partial\delta {\bf v}_s\over\partial t} = 
-\mbox{\boldmath$\nabla$}\delta\mu\label{eqfluid142}\eeq
\beq{\partial\delta s\over\partial t} 
+\mbox{\boldmath$\nabla$}\cdot(s_0\delta{\bf v}_n)=0\ 
..\label{eqfluid143}\eeq
Here the local equilibrium entropy density is $s(\br, t)=s_0+\delta 
s$, the local equilibrium pressure is $P(\br, t)=P_0+\delta P$ and 
the local chemical potential is $\mu(\br, t)=\mu_0+\delta\mu(\br, 
t).$ We note that the Landau hydrodynamic equations only involves 8 
equations, in contrast to the 9 equations we derived at the end of 
Section~\ref{sec:coupled}.  In particular, we see that Landau does 
not have {\em separate} continuity equations for the superfluid  
$\rho_s(\br, t)$ and normal fluid $\rho_n(\br, t)$ densities.   
Finally, it is assumed that these two components are always in local 
equilibrium with each other and hence the fluctuations are related by 
the thermodynamic identity
\beq n_0\delta\mu=-s_0\delta T+\delta P,\label{eqfluid144}\eeq
where $n_0$ is the equilibrium total density.

For simplicity, we now specialize our analysis to a uniform 
superfluid, where $\rho_{s0}, \rho_{n0}, s_0$ and other thermodynamic 
quantities are all position-independent.  Multiplying 
(\ref{eqfluid142}) by $n_{s0}$ gives [using (\ref{eqfluid144})]
\bea
\rho_{s0}{\partial\delta\bv_s\over\partial t} &=& 
-n_{s0}\mbox{\boldmath$\nabla$}\delta\mu\nonumber\\
&=&-{n_{s0}\over n_0}\mbox{\boldmath$\nabla$}(-s_0\delta T+\delta 
P)\nonumber\\
&=&-{n_{s0}\over n_0}\mbox{\boldmath$\nabla$}\delta P + {n_{s0}\over 
n_0}s_0\mbox{\boldmath$\nabla$}\delta T.\label{eqfluid145}\eea
Using this in (\ref{eqfluid138}) gives
\bea
-\mbox{\boldmath$\nabla$}\delta P &=& \rho_{s0} 
{\partial\delta\bv_s\over\partial t} 
+\rho_{n0}{\partial\delta\bv_n\over\partial t}\nonumber\\
&=&-{n_{s0}\over n_0}\mbox{\boldmath$\nabla$}\delta P+{n_{s0}\over 
n_0}s_0\mbox{\boldmath$\nabla$}\delta 
T+\rho_{n0}{\partial\bv_n\over\partial t}\nonumber\eea
or

\beq \rho_{n0}{\partial\delta\bv_n\over\partial t}=-{n_{n0}\over 
n_0}\mbox{\boldmath$\nabla$}\delta P-{n_{s0}\over 
n_0}s_0\mbox{\boldmath$\nabla$}\delta T.\label{eqfluid146}\eeq
Combining (\ref{eqfluid138}) and (\ref{eqfluid139}) gives
\beq{\partial^2\delta\rho\over\partial t^2} = 
-m\mbox{\boldmath$\nabla$}\cdot{\partial\delta{\bf j}\over\partial 
t}=\nabla^2\delta P.\label{eqfluid147}\eeq
Finally, (\ref{eqfluid143}) gives [using (\ref{eqfluid146})]
\bea {\partial^2\delta s\over\partial t^2} &=& - 
s_0\mbox{\boldmath$\nabla$}\cdot{\partial\delta{\bf v}_n\over\partial 
t}\nonumber \\
&=& {s_0\over\rho_0}\nabla^2\delta P 
+{s_0^2\over\rho_0}\left({\rho_{s0}\over\rho_{n0}}\right)\nabla^2\delta 
T\nonumber \\
&=& {s_0\over\rho_0}{\partial^2\delta\rho\over\partial t^2} 
+{s_0^2\over\rho_0}\left({\rho_{s0}\over\rho_{n0}}\right)\nabla^2\delta 
T.\label{eqfluid148}\eea
The last equation gives the local entropy fluctuations in terms of 
the local mass density $\delta\rho$ and temperature $\delta T$ 
fluctuations.
Eq.(\ref{eqfluid148}) can be re-written~\cite{Gorkov} in terms of the 
local entropy per unit mass $\bar s(\br, t)\equiv s(\br, t)/\rho(\br, 
t).$  Using
\beq \delta\bar s=-{s_0\over\rho_0^2}\delta\rho +{1\over\rho_0}\delta 
s,\label{eqfluid149}\eeq
(\ref{eqfluid148}) takes on the simpler form
\beq {\partial^2\delta\bar s\over\partial t^2} = \bar s_0^2 
\left({\rho_{n0}\over\rho_{s0}}\right)\nabla^2\delta T\ 
..\label{eqfluid150}\eeq
Expanding $\delta P$ and $\delta T$ in terms of $\delta\rho$ and 
$\delta\bar s$ fluctuations
\bea \partial P &=& \left.{\partial P\over\partial \rho}\right|_{\bar 
s}\delta\rho + 
\left.{\partial P\over\partial \bar s}\right|_{\rho}\delta\bar 
s\nonumber \\
\delta T &=&  \left.{\partial T\over\partial \rho}\right|_{\bar 
s}\delta\rho + 
\left.{\partial T\over\partial \bar s}\right|_{\rho}\delta\bar s\ 
,\label{eqfluid151}\eea
we see that (\ref{eqfluid150}) and (\ref{eqfluid147}) reduce to two 
coupled scalar equations for $\delta\rho$ and $\delta\bar s$.  
Inserting the normal mode solutions
\beq \delta\rho, \delta\bar s\propto e^{i({\bf k}\cdot\br -\omega 
t)},\label{eqfluid152}\eeq
one finds these coupled algebraic equations have two phonon solutions 
$\omega^2=u^2k^2$, where $u^2$ is the solution of the quadratic 
equation:
\bea u^4&-&u^2\left[\left.{\partial P\over\partial\rho}\right|_T 
+{T\over \bar c_v}\left.\left({1\over\rho_0}
{\partial P\over\partial 
T}\right|_\rho\right)^2+{\rho_{s0}\over\rho_{n0}} {T\bar s_0^2\over 
\bar c_v}\right]\nonumber \\
&+&{\rho_{s0}\over\rho_{n0}}\left({T\bar s_0^2\over\bar 
c_v}\right)\left.{\partial P\over\partial\rho}\right|_T = 
0.\label{eqfluid153}\eea
Here $\bar c_v = \left.{\partial \bar s\over\partial T}\right|_\rho$ 
is the equilibrium specific heat per unit mass.

The coefficients in (\ref{eqfluid153}) are daunting, but only involve 
equilibrium thermodynamic quantities which can be calculated for any 
given superfluid.  We emphasize that (\ref{eqfluid153}) is valid for 
both a uniform Bose-condensed gas and superfluid $^4$He, assuming 
that the superfluid and normal fluid are in local equilibrium with 
each other.  However, as we shall show, the detailed characteristics 
and behaviour of the two phonon modes (first and second second) are 
quite different in a Bose gas and in superfluid 
$^4$He.~\cite{Griffin97b,Leeyang}

A key feature about superfluid $^4$He is that (typical of any liquid) 
$\left.\partial P/\partial T\right|_\rho\simeq 0.$  This can be shown 
to be equivalent to $C_p\simeq C_v,$ where $C_{p,v}$ are the specific 
heats at constant pressure or constant volume.  In this case, 
(\ref{eqfluid153}) reduces to 
\beq u^4-u^2(A+B)+AB=0,\label{eqfluid154}\eeq
with two solutions $u^2=A, B:$
\bea
u^2_1 &=& {\partial P\over\partial\rho} \ \mbox{: first 
sound}\label{eqfluid155}\\
u^2_2 &=& {\rho_{s0}\over\rho_{n0}} \left({T\bar s_0^2\over\bar 
c_v}\right) \ \mbox{: second sound}.
\label{eqfluid156}\eea
Working out the associated motions, one finds the first sound mode 
$(\omega=u_1k)$ involves the in-phase motion of the superfluid and 
normal fluid components.  Moreover, one can show that first sound is 
essentially a pressure wave.  In contrast, the second sound mode 
$(\omega=u_2k)$ involves the out-of-phase motion of the two 
components, with $\delta{\bf j}\simeq 0$ or
\beq\rho_{n0}\,\delta\bv_n = -\rho_{s0}\,\delta \bv_s\ 
..\label{eqfluid157}\eeq
This corresponds to an almost pure temperature wave.  The successful 
detection in 1946 of a second sound mode (or temperature wave) was of 
tremendous significance in low temperature 
physics.~\cite{Gorkov,classic} The good agreement of the measured 
second sound velocity $u_2$ with the Landau expression in 
(\ref{eqfluid156}), calculated using the postulated phonon-roton 
quasiparticle spectrum, vindicated {\it both} aspects of the Landau 
theory of superfluid $^4$He.

In contrast, in a (Bose-condensed) gas, the pressure and temperature 
fluctuations are strongly coupled and hence $\partial P/\partial 
T|_\rho$ in (\ref{eqfluid153}) plays a significant role.  We recall 
that in a classical gas, one has $C_p/C_v={5\over 3}.$  Evaluating 
all the thermodynamic derivatives using our finite $T$ microscopic 
model (the single-particle HF spectrum in (\ref{eq:temperatures59})
replaces the roton spectrum of superfluid $^4$He) and working to 
lowest order in $gn_c/k_BT,$ one finds (after very lengthy 
calculations)
\beq u^2_1 = {5\over 3}{k_BT\over m}{g_{5/2}(z_0 = 1)\over 
g_{3/2}(z_0 = 1)}+O(gn)\label{eqfluid158}\eeq
\beq u^2_2 = {gn_{c0}\over m}+\dots\label{eqfluid159}\eeq
It turns out that the first sound mode in a Bose gas is largely an 
oscillation of the thermal cloud (the normal fluid). The second sound 
mode is largely an oscillation of the condensate (the superfluid).  
Both involve density fluctuations and thus will have significant 
weight in the dynamic structure factor related to the density 
response function.  The quite different features of first and second 
sound in Bose gases vs superfluid $^4$He should be remembered when 
reading standard texts about superfluid $^4$He. Results equivalent to 
(\ref{eqfluid158}) and (\ref{eqfluid159}) are given in 
Refs.~\cite{Griffin97b,Leeyang,Shenoy,Kirkdorfman}.

Comparing (\ref{eqfluid159}) with (\ref{eq:dynamics28}), it is clear 
that in a uniform Bose-condensed gas, the {\it hydrodynamic} second 
sound mode at finite $T$ smoothly extrapolates to the Bogoliubov 
phonon mode in the {\it collisionless} region.  The fact that the 
first sound velocity $u_1$ does not depend on the interaction 
strength $g$ to lowest order is typical of ordinary sound waves in 
any gas.  However, one must remember that interactions (collisons) 
play a crucial indirect role in enforcing dynamic local equilibrium.

We now turn to a discussion of the generalized two-fluid hydrodynamic 
equations we derived at the end of Section~\ref{sec:coupled}.  Again, 
for simplicity, we limit our analysis to a uniform Bose-condensed 
gas. If we can ignore vorticity in both fluids, we can introduce  two 
velocity potentials
\bea
\delta\bv_c &\equiv&\mbox{\boldmath$\nabla$}\phi_c(\br, t)\nonumber\\
\delta\bv_n &\equiv&\mbox{\boldmath$\nabla$}\phi_n(\br, 
t),\label{eqfluid160}\eea
and reduce the equations in (\ref{eqcoupled134}) and 
(\ref{eqcoupled135}) to three equations for $\phi_c, \phi_n$ and 
$\delta\mu_{diff}$ [the latter is defined in (\ref{eqcoupled133})]:
\bea
m{\partial^2\phi_c\over\partial t^2} &=& gn_{c0}\nabla^2\phi_c 
+2g{\tilde 
n}_0\nabla^2\phi_n+{\sigma_H\over\tau_\mu}\delta\mu_{diff}
\label{eqfluid161}\\
m{\partial^2\phi_n\over\partial t^2} &=& \left({5\over 3}{{\tilde 
P}_0\over{\tilde n}_0}+2g{\tilde 
n}_0\right)\nabla^2\phi_n+2gn_{c0}\nabla^2\phi_c -{2\over 
3}{\sigma_H\over\tau_\mu}\delta\mu_{diff}
\label{eqfluid162}\eea
\beq{\partial\delta\mu_{diff}\over\partial t}={2\over 
3}gn_{c0}\nabla^2\phi_n-gn_{c0}\nabla^2\phi_c-
{\delta\mu_{diff}\over\tau_\mu}.\label{eqfluid163}\eeq
Here $\tau_\mu$ is a new relaxation time governing how 
$\delta\mu_{diff}$ relaxes to $0$ (ie, ${\tilde \mu}\to\mu_c)$.  It 
is found to be related to the $C_{12}$ collision time $\tau^\prime$ 
defined in 
(\ref{eq:temperatures80}) by
\beq{1\over\tau_\mu} = 
{1\over\sigma_H\tau^\prime},\label{eqfluid164}\eeq
where the dimensionless hydrodynamic renormalization factor 
$\sigma_H$ is given by~\cite{Zarnikgrif,Nikuni99}\beq \sigma_H \equiv 
{{5\over 2}{\tilde\gamma}_0{\tilde P}_0 -{3\over 2} g{\tilde 
n}_0^2\over
{5\over 2}{\tilde P}_0(1-{\tilde\gamma}_0)+2g{\tilde n}_0 
n_{c0}+{2\over 3}g{\tilde\gamma}_0 n^2_{c0} +{3\over 2} g{\tilde 
n}^2_0}\ .\label{eqfluid165}\eeq
For completeness, we recall that 
\bea
{\tilde P}_0 (z_0) &=& {k_BT\over\Lambda^3_0}g_{5/2}(z_0= e^{-\beta 
gn_{c0}})\ \nonumber \\
{\tilde n}_0 (z_0) &=& {1\over\Lambda^3_0} g_{3/2} (z_0)\nonumber \\
{\tilde\gamma}_0 &=& {g\over k_BT\Lambda_0^3} 
g_{1/2}(z_0),\label{eqfluid166}\eea
where the Bose-Einstein functions are $g_n(z)\equiv\sum^\infty_{l=1} 
z^l/l^n$ and $\Lambda_0$ is the equilibrium thermal de Broglie 
wavelength.  Calculation shows that $\sigma_H$ becomes large as $T\to 
T_{BEC}$ and thus $\tau_\mu$ can become very large near the 
superfluid transition.  This is not in contradiction with the fact 
that the present discussion is for the collision-dominated region 
$[\omega\tau_{22}\ll 1,$ where $\tau_{22}$ is some relaxation time 
associated with the $C_{22}$ collision integral in 
(\ref{eqcoupled88})].
\begin{figure}
\vspace{-0.9in}
  \centerline{\epsfig{file=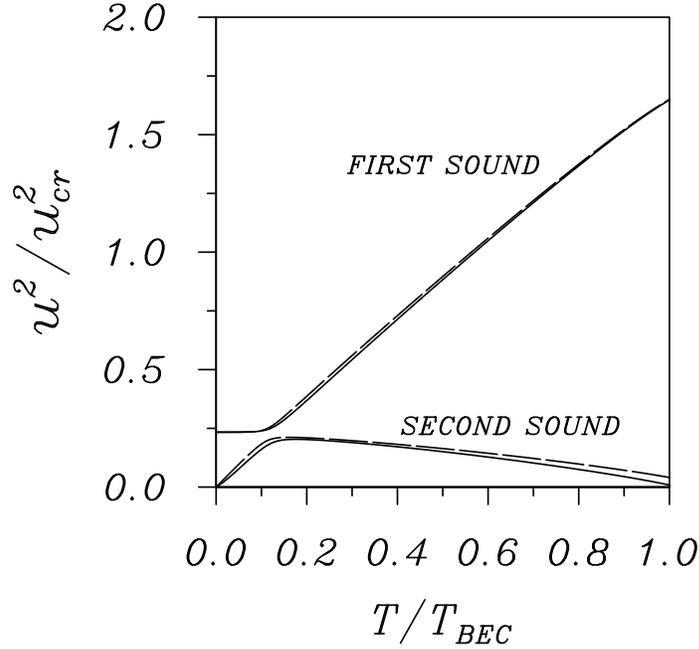,width=4.4in}}
\vspace{-1.4in}
\caption{Squares of the first and second sound velocities (normalized 
by the
first sound velocity of the ideal gas at $T = T_{BEC}$)
vs. $T/T_{BEC}$. The solid lines are for the Landau limit $\omega
\tau_\mu \ll 1$ while the dashed lines are for 
$\omega \tau_\mu \gg 1$. The curves were generated for $gn/k_B T_{BEC}
= 0.2$. The interesting crossover at low temperatures was first 
pointed out in Ref.~\protect\cite{Leeyang}. It is reproduced by our 
equations, even though it occurs when $gn_{c0}\sim k_{B}T$ which is 
outside the region of validity of our model. See Refs.~\protect\cite{Zarnikgrif,Griffin97b}.} \label{fig:canberra5}
\end{figure}

It is clear from the fact that we have 3 coupled equations for the 3 
variables $\phi_c, \phi_n$ and $\delta\mu_{diff}$, we will obtain a 
new mode, in addition to the usual first and second sound 
oscillations discussed earlier.  It is convenient to first eliminate 
$\delta\mu_{diff}$ using the solution of (\ref{eqfluid163}),
\begin{displaymath}
-i\omega\delta\mu_{diff} = {2\over 
3}gn_{c0}(-k^2)\phi_n-gn_{c0}(-k^2)\phi_c-{\delta\mu_{diff}
\over\tau_\mu}\end{displaymath}

or
\beq \delta\mu_{diff} = {gn_{c0}\tau_\mu\over 1-i\omega\tau_\mu} 
(\phi_c-{2\over 3}\phi_n)k^2.\label{eqfluid167}\eeq
Using this in (\ref{eqfluid161}) and (\ref{eqfluid162}) gives
\beq m\omega^2\phi_c = gn_{c0}\left(1-{\sigma_H\over 
1-i\omega\tau_\mu}\right) k^2\phi_c +
2g{\tilde n}_0\left(1+{\sigma_H\over 3(1-i\omega\tau_\mu)} 
{n_{c0}\over{\tilde n}_0}\right)k^2\phi_n\label{eqfluid168}\eeq
\bea
m\omega^2\phi_n &=& \left({5\over 3}{{\tilde P}_0\over{\tilde n}_0} + 
2g{\tilde n}_0\left[1-{2\sigma_H\over 9(1-i\omega\tau_\mu)} 
{n^2_{c0}\over{\tilde n}_0^2}\right]\right)k^2\phi_n\nonumber\\
&+& 2gn_{c0} \left(1+{\sigma_H\over 3(1-i\omega\tau_\mu)} 
{n_{c0}\over{\tilde n}_0}\right)k^2\phi_c\ .\label{eqfluid169}\eea
These two coupled equations for $\phi_n$ and $\phi_c$ are easily 
solved and we obtain (as expected) first and second sound modes.  
Clearly the velocities $u_1$ and $u_2$ will now depend on the value 
of $\omega\tau_\mu,$ although it turns out that this dependence is 
not very strong.
This is shown by the results in Fig.~\ref{fig:canberra5}.

The $\omega\tau_\mu\to 0$ limit is of special interest since the 
equations (\ref{eqfluid168}) and (\ref{eqfluid169}) can then be shown 
to be completely equivalent to the predictions of the Landau 
two-fluid equations discussed earlier in this section.  To be 
precise, the first and second sound velocities $u_{1,2}^2$ obtained 
from (\ref{eqfluid168}) and (\ref{eqfluid169})  in the limit 
$\omega\tau_\mu\to 0$ agree with those given by (\ref{eqfluid153}).  
This equivalence makes physical sense since in the limit of 
$\tau_\mu\to 0,$ one sees that $\delta\mu_{diff}\to 0$ very rapidly 
[see (\ref{eqfluid163}) and (\ref{eqfluid167})].  Thus we have proven 
that the Landau two-fluid hydrodynamic equations are correct if the 
condensate and thermal cloud are in diffusive local equilibrium.  
This, as noted earlier, is an (sometimes implicit) assumption in all 
previous derivations of the Landau equations.~\cite{Bogg,Kirkdorfman}

\begin{figure}
\centerline{\epsfig{file=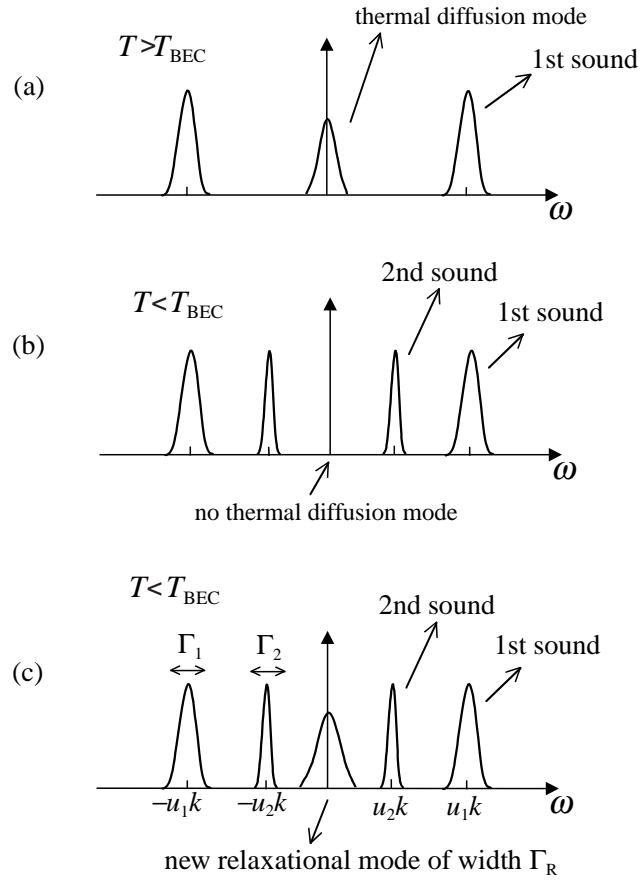,width=4.4in}}
\vspace{-0.8in}
\caption{A schematic illustration  of the predicted hydrodynamic mode 
spectrum
for a uniform Bose gas.  (a) Above $T_{\rm BEC}$, one has the usual
first sound mode and a zero-frequency thermal diffusion mode.  (b)
Below $T_{\rm BEC}$, the standard Landau-Khalatnikov two-fluid
hydrodynamics predicts first and second sound, but no remnant of any
zero-frequency mode.  (c) In the ZGN$'$ theory presented in Sections 4
and 5, one has first and second sound modes as well as a
zero-frequency relaxational mode.  Based on 
Ref.~\protect\cite{Nigriza}.} \label{fig:canberra6}
\end{figure}

This Landau limit $(\omega\tau_\mu\to 0)$ is, in fact, very subtle in 
the context of our microscopic calculation.  We see that in this 
limit, there are still correction terms in (\ref{eqfluid168}) and 
(\ref{eqfluid169}) which are proportional to the hydrodynamic 
renormalization factor $\sigma_H$  as defined in (\ref{eqfluid165}).  
These terms are crucial in ensuring that (\ref{eqfluid168}) and 
(\ref{eqfluid169}) reproduce the results of the Landau two-fluid 
equations.  Another way of seeing this is that even though 
$\delta\mu_{diff}\to 0$ when $\tau_\mu\to 0,$ we note that  
$\delta\Gamma_{12}$ in (\ref{eqcoupled132}) is still finite.
Using (\ref{eqfluid167}), one finds
\beq \delta\Gamma_{12} =-n_{c0}\sigma_H\left(\phi_c -{2\over 
3}\phi_n\right)k^2.\label{eqfluid170}\eeq
In conclusion, one might say that the hydrodynamic renormalization 
factor $\sigma_H$ contains a key part of the physics buried in the 
hydrodynamic two-fluid equations of Landau.

In the opposite limit $\omega\tau_\mu\gg 1$ (which can arise near 
$T_{BEC}$) we see all the terms proportional to $\sigma_H$ in 
(\ref{eqfluid168}) and (\ref{eqfluid169}) are negligible.  This 
domain is missed in the Landau two-fluid equations.  It describes 
situations in which the condensate and thermal cloud are out of 
diffusive equilibrium with each other.  It would be of great interest 
to look for this kind of phenomenon in trapped Bose gases.

One can work out the frequency of the new mode associated with the 
dynamics of $\delta\mu_{diff}$ and for a uniform gas, it is well 
approximated by 
\beq \omega_R\simeq -i/\tau_\mu.\label{eqfluid171}\eeq
Thus, in general, our two-fluid hydrodynamic equations predict the 
existence of a relaxational mode peaked at zero frequency.  One can 
improve the theory to include deviations from local equilibrium [ie, 
deviations of $f$ from ${\tilde f}$ in (\ref{eqcoupled116})].  This 
involves a Chapman-Enskog kind of calculation familiar in the theory 
of classical gases and gives rise to hydrodynamic 
damping.~\cite{Statistical}  The normal fluid (non-condensate) 
equations of motion have new terms corresponding to shear viscosity 
$(\eta)$ and thermal conductivity $(\kappa)$ transport 
coefficients~\cite{Kirkdorfman,Nigriza}, where we recall that 
$\kappa$ and $\eta$ are proportional to some $\tau_{22}$ collision 
time and hence go as $1/g^2$.  The damping of first sound, second 
sound and the relaxational mode due to small deviations from local 
equilibrium among the thermal atoms $(C_{22}[f, \Phi])$ has been 
recently worked out in detail by Nikuni, Zaremba and the 
author.~\cite{Nigriza}  Of particular interest is the effect on the 
relaxational mode, which is now described by [compare with 
(\ref{eqfluid171})]  
\beq \omega_R \simeq -i\left[{1\over\tau_\mu} +A\kappa 
k^2\right].\label{eqfluid172}\eeq
Effectively the relaxation mode is strongly coupled into thermal 
conduction processes (or flow of heat).  The expression for the 
coefficient $A$ in (\ref{eqfluid172}) is somewhat 
complicated~\cite{Nigriza} but above $T_{\rm BEC}$  
$({1\over\tau_\mu} \to 0),$ we find the mode reduces to the 
well-known thermal diffusion mode,
\beq \omega_R = -i{\kappa k^2\over n_0 C_p} = -i D_T k^2\ 
..\label{eqfluid173}\eeq
This result strongly suggests that our new relaxational mode below 
$T_{\rm BEC}$ is the renormalized version of the usual thermal 
diffusion mode above $T_{\rm BEC}$.

We recall that in the standard two-fluid hydrodynamic theory of 
Landau, the thermal diffusion mode below $T_{\rm BEC}$ disappears 
below $T_{\rm BEC}$, with, its spectral weight going into the 
emerging second sound doublet.  In Fig.~\ref{fig:canberra6}, we 
schematically illustrate the different hydrodynamic mode spectra 
predicted for above and below $T_{\rm BEC}$.

To illustrate the essential physics, the detailed discussion in this 
section has been limited to a uniform Bose-condensed gas.  The 
analysis can, of course, be extended to trapped Bose gas.  In 
particular, the out-of-phase dipole 
mode~\cite{Zargrifnikun,Zarnikgrif} in the hydrodynamic limit is of 
special interest.  This corresponds to centre-of-mass oscillation of 
the equilibrium condensate and non-condensate density profiles,
\bea n_c(\br, t) &=& n_{c0} (\br - 
\mbox{\boldmath$\eta$}_c(t))\nonumber\\
{\tilde n}(\br, t) &=& {\tilde n}_0 (\br - 
\mbox{\boldmath$\eta$}_n(t)),\label{eqfluid174}\eea
with \beq\bv_n = {d\mbox{\boldmath$\eta$}_n\over dt} \ ; \ \bv_c = 
{d\mbox{\boldmath$\eta$}_c\over dt}\ .\label{eqfluid175}\eeq
One finds that the hydrodynamic equations give (as expected) an 
undamped {\em in-phase} normal mode, with $\bv_n = \bv_c$ and 
$\omega=\omega_0$, where $\omega_0$ is the trap frequency.  This is a 
generalized version of the $T = 0$ Kohn mode discussed at the end of 
Section~\ref{sec:Pure}.  In addition, however, there is an {\em 
out-of-phase} mode satisfying
\beq N_c\bv_c = -{\tilde N}\bv_n,\label{eqfluid176}\eeq
\begin{figure}
\vspace{-1.4in}
  \centerline{\epsfig{file=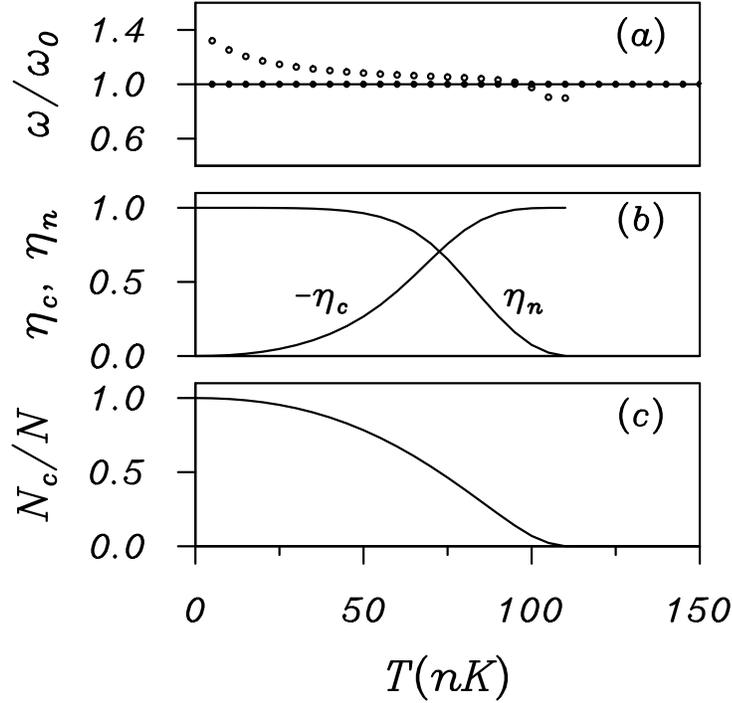,width=4.4in}}
\vspace{-0.6in}
\caption{(a) Mode frequencies for the in-phase (solid dots) and
out-of-phase (open dots) dipole modes vs. temperature
for 2000 Rb atoms in an isotropic parabolic trap (see
Ref. 13 for values of the physical parameters used).
(b) Condensate ($\eta_c$) and non-condensate ($\eta_n$) 
amplitudes for the out-of-phase dipole mode.
(c) Fraction of atoms in the condensate as a function of
temperature.  From Ref.~\protect\cite{Zargrifnikun}.}
\label{fig:canberra7}
\end{figure}
\noindent with a frequency different from the trap frequency 
$\omega_0$.  The frequency of this mode is shown in 
Fig.~\ref{fig:canberra7}. Calculation shows that this {\it 
out-of-phase} mode is only damped by its coupling to the relaxational 
mode given by (\ref{eqfluid171}).  There is no hydrodynamic-type 
damping from the finite transport coefficients of the normal gas.  
Calculations are in reasonable agreement with the observed damping of 
such a out-of-phase mode.~\cite{Stamper-Kurn98} A careful study of 
this out-of-phase dipole mode as a function of the temperature would 
be a nice way of probing the unusual hydrodynamics of a 
Bose-condensed gas.

In these lectures, I have not made any attempt to analyze the 
available experimental 
investigations~\cite{cornell,ketterle,Jin97,Stamper-Kurn98} of 
collective modes at finite temperatures.  These pioneering 
experimental studies are very promising but we need more systematic 
investigations, especially as a function of temperature and density.  
A key reason why such studies are perhaps a unique way of probing the 
many-body dynamics of superfluid gases~\cite{Pitstring} is that one 
can measure collective mode frequencies very accurately (at $T=0$, 
with errors of only a few percent).

In this section, we have put emphasis on deriving two-fluid 
hydrodynamic equations by starting from an approximate but still 
microscopic model.  Such a derivation has been carried out recently 
by Zaremba, Nikuni and the author~\cite{Zarnikgrif} for a trapped 
Bose gas.  The pioneering work on deriving two-fluid hydrodynamic 
equations for a uniform dilute Bose gas was by Kirkpatrick and 
Dorfman.~\cite{Kirkdorfman} In this regard, it is useful to point out 
that the linearized Landau two-fluid hydrodynamic equations [as given 
by (\ref{eqfluid138}) - (\ref{eqfluid143})] are expected to be {\it 
exact} as long as the two components are in local equilibrium.  Thus, 
for a uniform system, the exact first and second sound velocities are 
given by the solutions of (\ref{eqfluid153}).  The only question is 
how to calculate the various thermodynamic functions which are 
involved in this equation.  The analogous Landau equations for a 
trapped gas allows us to go past the simplified microscopic models we 
have used in Section~{\ref{sec:coupled} to derive equations of this 
kind. In particular, one should be able to use the Landau two-fluid 
equations as a direct probe of the superfluid density, just as one 
does in superfluid $^4$He.~\cite{Gorkov,classic} One might be able to 
detect small differences between the magnitude of the superfluid 
density $n_{s0}$ and the condensate density $n_{c0}$ (at the level of 
our model in Section~{\ref{sec:coupled}, $n_{s0}$ and $n_{c0}$ are 
equal). In this connection, we note that in the formal zero 
temperature limit, the Landau two-fluid equations reduce to two 
coupled equations for a {\em pure} superfluid $(\rho_s =\rho, 
\rho_n=0$ at $T=0)$.  This limit has been used~\cite{Pitstring} to 
find a more accurate version of the $T=0$ quantum hydrodynamic 
equations discussed in Section~{\ref{sec:Pure}.

In conclusion, I hope I have given some insight into why the 
two-fluid hydrodynamics of trapped Bose gases has so much potential 
interest.  I hope some of the young experimentalists attending this 
Summer School will take up the challenge to study this new frontier. 
In a sense, the natural next step after understanding the dynamics of 
a pure condensate $(T=0)$ is to study the two-fluid hydrodynamics of 
trapped gases (at finite $T$) since now one has {\em two} components 
which can execute coupled coherent collective motions.  In contrast, 
the collisionless region at finite $T$ as discussed in 
Section~{\ref{sec:temperatures} does not seem as interesting since 
the thermal cloud has always such a low density (see 
Fig.~\ref{fig:canberra2}).  As a result, mean field effects produced 
by the non-condensate are never very 
important~\cite{Kavoulakis98,Guery}, above or below $T_{\rm BEC}$, 
and thus no new many-body dynamics emerges which is different than 
already found at $T=0$.  What is new is the different mechanisms of 
damping of collisionless condensate motion, as discussed in 
Sections~\ref{sec:temperatures} and \ref{sec:coupled}.

\section*{Acknowledgments}
I am grateful to Craig Savage and Mukunda Das for organizing this 
Australian BEC Summer School, for inviting me to participate, and 
their hospitality and assistance. Much of the work reviewed in these 
lectures was carried out with Eugene Zaremba, Tetsuro Nikuni, Jamie 
Williams and Milena Imamovi\'c-Tomasovi\'c.  I would like to thank 
Jamie Williams for a critical reading and help with preparing the 
manuscript.  I also thank Helen Iyer for texing the manuscript in 
record time.  My research is supported by NSERC of Canada.

 \end{document}